\documentclass[twocolumn,tighten,twocolappendix]{aastex63}

\usepackage{hyperref}
\usepackage{subfigure}
\usepackage{booktabs}
\usepackage{soul}

\newcommand{\oii}{\ion{O}{2}}

\newcommand{\nvi}{\ion{N}{6}}
\newcommand{\nvii}{\ion{N}{7}}
\newcommand{\ovii}{\ion{O}{7}}
\newcommand{\oviii}{\ion{O}{8}}
\newcommand{\neix}{\ion{Ne}{9}}
\newcommand{\nex}{\ion{Ne}{10}}
\newcommand{\nvin}{\ion{N}{6}}
\newcommand{\nviin}{\ion{N}{7}}
\newcommand{\oviin}{\ion{O}{7}}
\newcommand{\oviiin}{\ion{O}{8}}
\newcommand{\neixn}{\ion{Ne}{9}}
\newcommand{\nexn}{\ion{Ne}{10}}
\newcommand{\mgxi}{\ion{Mg}{11}}
\newcommand{\mgxii}{\ion{Mg}{12}}
\newcommand{\fexvin}{\ion{Fe}{16}}
\newcommand{\fexviin}{\ion{Fe}{17}}
\newcommand{\fexviiin}{\ion{Fe}{18}}
\newcommand{\fexvii}{\ion{Fe}{17}}
\newcommand{\fexxi}{\ion{Fe}{21}}
\newcommand{\fexxii}{\ion{Fe}{22}}
\newcommand{\fexxv}{\ion{Fe}{25}}

\def\cii{{{\rm C}\,{\sc ii}~}}
\def\ciin{{{\rm C}\,{\sc ii}}}
\def\civ{{{\rm C}\,{\sc iv}~}}
\def\civn{{{\rm C}\,{\sc iv}}}
\def\cv{{{\rm C}\,{\sc v}~}}
\def\cvn{{{\rm C}\,{\sc v}}}
\def\cvi{{{\rm C}\,{\sc vi}~}}
\def\cvin{{{\rm C}\,{\sc vi}}}
\def\ni{{{\rm N}\,{\sc i}~}}
\def\nin{{{\rm N}\,{\sc i}}}
\def\nii{{{\rm N}\,{\sc ii}~}}
\def\niii{{{\rm N}\,{\sc iii}~}}
\def\nv{{{\rm N}\,{\sc v}~}}
\def\nvi{{{\rm N}\,{\sc vi}~}}
\def\nvii{{{\rm N}\,{\sc vii}~}}

\def\oii{{{\rm O}\,{\sc ii}}}

\def\oiv{{{\rm O}\,{\sc iv}~}}
\def\ov{{{\rm O}\,{\sc v}~}}
\def\ovn{{{\rm O}\,{\sc v}}}
\def\ovi{{{\rm O}\,{\sc vi}~}}
\def\ovin{{{\rm O}\,{\sc vi}}}
\def\ovii{{{\rm O}\,{\sc vii}~}}
\def\oviii{{{\rm O}\,{\sc viii}~}}
\def\neviiin{{{\rm Ne}\,{\sc viii}}}
\def\neix{{{\rm Ne}\,{\sc ix}~}}
\def\nex{{{\rm Ne}\,{\sc x}~}}
\def\mgxi{{{\rm Mg}\,{\sc xi}~}}
\def\mgxin{{{\rm Mg}\,{\sc xi}}}
\def\mgxii{{{\rm Mg}\,{\sc xii}~}}
\def\sixiii{{{\rm Si}\,{\sc xiii}~}}
\def\sixiv{{{\rm Si}\,{\sc xiv}~}}
\def\sixivn{{{\rm Si}\,{\sc xiv}}}
\def\nvin{{{\rm N}\,{\sc vi}}}
\def\nviin{{{\rm N}\,{\sc vii}}}
\def\oviin{{{\rm O}\,{\sc vii}}}
\def\oviiin{{{\rm O}\,{\sc viii}}}

\def\neixn{{{\rm Ne}\,{\sc ix}}}
\def\nexn{{{\rm Ne}\,{\sc x}}}
\def\fexvin{{{\rm Fe}\,{\sc xvi}}}
\def\fexviin{{{\rm Fe}\,{\sc xvii}}}
\def\fexvii{{{\rm Fe}\,{\sc xvii}~}}
\def\fexix{{{\rm Fe}\,{\sc xix}~}}
\def\fexxi{{{\rm Fe}\,{\sc xxi}~}}
\def\fexviiin{{{\rm Fe}\,{\sc xviii}}}
\def\fexxii{{{\rm Fe}\,{\sc xxii}}}
\def\fexxv{{{\rm Fe}\,{\sc xxv}~}}
\def\nixxiv{{{\rm Ni}\,{\sc xxiv}~}}

\def\xmm{{\it XMM-Newton}~}
\def\chandra{{\it Chandra}~}

\def\msun{{M$_{\odot}$}}

\shorttitle{complex phase structure of 10$^{5-8}$\,K Galactic halo}
\shortauthors{Das et al.}
\begin{document}

\title{The hot circumgalactic medium of the Milky Way: 
evidence for super-virial, virial, and sub-virial temperature, non-solar chemical composition, and non-thermal line broadening
}
\correspondingauthor{Sanskriti Das}
\email{das.244@buckeyemail.osu.edu}

\author[0000-0002-9069-7061]{Sanskriti Das}
\affiliation{Department of Astronomy, The Ohio State University, 140 West 18th Avenue, Columbus, OH 43210, USA}

\author[0000-0002-4822-3559]{Smita Mathur}
\affiliation{Department of Astronomy, The Ohio State University, 140 West 18th Avenue, Columbus, OH 43210, USA}
\affil{Center for Cosmology and Astroparticle Physics, 191 West Woodruff Avenue, Columbus, OH 43210, USA}
\author[0000-0003-1880-1474]{Anjali Gupta}
\affiliation{Department of Astronomy, The Ohio State University, 140 West 18th Avenue, Columbus, OH 43210, USA}
\affil{Columbus State Community College, 550 E Spring St., Columbus, OH 43210, USA}
\author[0000-0001-6291-5239]{Yair Krongold}
\affiliation{Instituto de Astronomia, Universidad Nacional Autonoma de Mexico, 04510 Mexico City, Mexico}

\begin{abstract}
\noindent For the first time, we present the simultaneous detection and characterization of three distinct phases at $>10^5$\,K in $z=0$ absorption, using deep \chandra observations toward Mrk\,421. The extraordinarily high signal-to-noise ratio ($\geqslant60$) of the spectra has allowed us to detect a \textit{hot} phase of the Milky\,Way circumgalactic medium (CGM) at 3.2$^{+1.5}_{-0.5}\times$ 10$^7$\,K, coexisting with a \textit{warm-hot} phase at 1.5$\pm$0.1$\times$10$^6$\,K and a \textit{warm} phase at 3.0$\pm$0.4$\times$10$^5$\,K. The \textit{warm-hot} phase is at the virial temperature of the Galaxy, and the \textit{warm} phase may have cooled from the \textit{warm-hot} phase, but the super-virial \textit{hot} phase remains a mystery. We find that {[C/O]} in the \textit{warm} and \textit{warm-hot} phases, [Mg/O] in the \textit{warm-hot} phase and [Ne/O] in the \textit{hot} phase are super-solar, and the \textit{hot} and the \textit{warm-hot} phases are $\alpha-$enhanced. Non-thermal line broadening is evident in the \textit{warm-hot} and the \textit{hot} phases and it dominates the total line broadening. {Our results indicate that the $>10^5$\,K CGM is a complex ecosystem.
It provides insights on the thermal and chemical history of the Milky\,Way CGM, and 
theories of galaxy evolution. }
\end{abstract}

\section{Introduction} \label{sec:intro}
\noindent The circumgalactic medium (CGM) is the halo of multi-phase gas and dust surrounding the stellar component and interstellar medium (ISM) of galaxies, extended out to their virial radii \citep{Tumlinson2017}\footnote{The circumgalactic gas of the Milky Way (MW) is usually referred as the Galactic ``halo" or ``corona". CGM is a more prevalent term for external galaxies. Because they all mean essentially the same, we will use these terms interchangeably.}. It plays an instrumental role in the evolution of a galaxy as a nexus of the accretion, feedback and recycling \citep{Putman2012,Voit2015,Schaye2015,Nelson2018} by harboring a large fraction of its missing baryons and missing metals \citep{Gupta2012,Peeples2014}. The temperature of the CGM is predicted to span at least two orders of magnitude: T $\approx$ 10$^{4-6}$\,K, but most of its mass is \textit{believed} to reside in the volume-filling warm-hot (T $\gtrapprox$ 10$^6$\,K) component \citep{Oppenheimer2016,Nelson2018,LiMiao2017}\footnote{Traditionally, ``warm-hot" refers to the temperature range of T$=10^5$--$10^7$\,K \citep{Cen1999}. In this paper, we refer to the $10^6$--$10^7$\,K range as warm-hot and the $10^5$--$10^6$\,K range as warm}. 

The warm-hot gaseous Galactic corona at the virial temperature has been predicted for a long time \citep{Spitzer1956}. For $\geqslant$ 10$^{12}$ M$_\odot$ halos, the warm-hot CGM is expected to be an amalgam of the rarefied, metal-poor, shock-heated infalling gas that has not yet cooled and fallen to the disk, and the dense metal-enriched galactic outflow driven by the winds of massive stars, supernovae and AGN (active galactic nuclei) feedback. Therefore, the warm-hot CGM is not necessarily a phase at a single-temperature and of solar-like chemical composition. Studying the abundances in the highly ionized CGM and its different thermal components, if any, is extremely important to understand the thermal and chemical evolution of the CGM and any contribution from the sources of non-thermal energy. Deep X-ray absorption spectroscopy, where the warm-hot CGM can be probed by He-like and H-like ionized metals, provides a great tool. 

Because of our special vantage point, the warm-hot CGM of the Milky\,Way has been studied via emission and absorption in much better details compared to other galaxies. The combined studies of emission and absorption show that the warm-hot CGM is at T $\approx$ 10$^{6.3}$\,K, and it is diffuse, extended, massive, and anisotrpic \citep{Kuntz2000,Gupta2009,Henley2010,Gupta2012,Henley2013,Gupta2014,Nicastro2016b,Gupta2017,Gatuzz2018}. From both the emission and absorption studies focused on \ovii and \oviii lines, the warm-hot CGM was found to be consistent with a single temperature. There were hints of hotter components in some emission studies, but these were questionable due to confusion with the foreground components \citep{Yoshino2009,Henley2013,Nakashima2018}.  

By analyzing a deep spectrum toward the blazar 1ES\,1553+113 observed with \xmm Reflection Grating Spectrometer (exposure time = 1.85\,Ms), \citet{Das2019a} discovered a $\approx10^7$\,K CGM component coexisting with the $\approx10^6$\,K component. The discovery of the $\approx10^7$\,K hot component, driven by the detection of the \nex line, was unambiguous and robust. The emission analysis around the same sightline revealed that the temperatures of the emitting and the absorbing gas were not the same \citep{Das2019c}, unlike the previous studies of one-temperature halo model \citep{Gupta2012,Gupta2014,Gupta2017}. This showed that the highly ionized halo gas consists of at least three components, and the picture of an isothermal CGM at the virial temperature is clearly ruled out. The evidence of the hotter CGM component(s) has further been supported by emission analysis toward other sightlines \citep{Gupta2021}. 

The discovery of the $\approx10^7$\,K gas at super-virial temperature in the Milky\,Way CGM was surprising and perplexing. How ubiquitous is the $\approx10^7$\,K hot gas? Detecting the weak lines probing the hot component requires exceedingly high signal-to-noise ratio (S/N) spectra. In this paper, we present analysis of the $z=0$ absorbers toward the blazar Mrk\,421 using the archival \chandra data. The sightline probes the Milky\,Way CGM in absorption with unprecedented sensitivity. Detailed spectral analysis and ionization modeling of the data have led to the discovery of a very \textit{hot} component of the Milky\,Way CGM at T$_3$ $\approx10^{7.5}$\,K, coexisting with two phases at T$_1$ $\approx 10^{5.5}$\,K and T$_2$ $\approx 10^{6.2}$\,K. Non-thermal broadening is found to dominate the total line broadening in the two hotter phases. We also find that the phases at T$_2$ and T$_3$ are $\alpha-$enhanced, indicating core-collapse supernovae enrichment. Additionally, we find non-solar abundance ratios of $\alpha$-elements (C, O, Ne, Mg and Si). 

Our paper is structured as follows. We discuss data reduction and analysis in section \ref{sec:analysis}, and results in section \ref{sec:results}. We interpret our results and discuss their implications in section \ref{sec:discussion}. Finally, we summarize our results and outline the future aim in section \ref{sec:summary}.

\section{Data reduction and analysis}\label{sec:analysis}
\noindent Mrk\,421 ($l, b$ = 179.83$^\circ$, 65.03$^\circ$) is one of the nearby ($z=0.031$) brightest blazars in the sky. For the past 20 years, it has been observed by multiple detectors and gratings of \chandra both as a target of opportunity (ToO) and as a calibration source. Here, we focus on the $z=0$ absorption lines of multiple metals in the 4.9--43.4 \AA~range of the spectra, probing the circumgalactic medium (CGM) of the Milky\,Way. We use the optimum combination of the detectors and gratings according to their effective area and spectral resolution. With the medium energy grating (MEG) arm of the high-energy transmission grating (HETG) of ACIS-S, we probe the magnesium (Mg) and silicon (Si) lines. With the low-energy transmission grating (LETG) of ACIS-S we probe neon (Ne) and oxygen (O) lines. And with HRC-S/LETG we probe nitrogen (N) and carbon (C) lines. Thus, the three instrument-setting together allows us to probe the large wavelength range we report here, which is critical to search for multiple temperature components.

We use {$\chi^2$ (\texttt{XSPEC} command} \texttt{chi}) statistic throughout our analysis. In the following sections, we quote 1$\sigma$ error bars and 68\% confidence intervals, unless explicitly mentioned otherwise. 

\subsection{Data extraction and reduction}
\noindent From the public archive of \chandra we extract all the archival data of ACIS-MEG\footnote{ObsID: 1714, 10663, 10670, 13098, 13105, 14320, 14327, 15477, 15484, 16424, 16431, 17385, 17392, 19853, 19860, 20710, 20717, 21816, 21823} and HRC-LETG\footnote{ObsID: 1715, 4149, 8396, 10665, 10667, 10669, 12122, 13100, 13102, 13104, 14322, 14323, 14324, 14326, 14396, 14397, 15479, 15481, 15483, 16426, 16428, 16430, 17387, 17389, 17391, 19444, 19855, 19857, 19859, 20712, 20714, 20716, 20942, 20943, 20944, 21818, 21820, 21822}. The effective area of ACIS-LETG degraded significantly after cycle 15, therefore we use the data only before then\footnote{ObsID: 4148, 5171, 5318, 5332, 6925, 8378, 10664, 10668, 10671, 11966, 11970, 11972, 11974, 12121, 13097, 13099, 13101, 13103, 14266, 14321, 14325, 15476, 15478, 15480, 15482, 15607}. 

We reprocess and reduce all the data using the latest calibration database of CIAO\,4.13 with the \texttt{chandra\_repro} command. It produces the spectra, background spectra, response matrices (RMF) and auxiliary response files (ARF) for each observation ID. We combine the positive and negative 1st order spectra of all observation IDs separately for ACIS-MEG, ACIS-LETG and HRC-LETG, using the  \texttt{combine\_grating\_spectra} command. In HRC-LETG, the diffraction orders of the spectra cannot be resolved, so the spectrum is a combination of all orders. To account for the contribution of higher orders in the spectra, we combine the 2nd to 8th order positive and negative RMFs and ARFs, separately for each order, of each observation ID using the \texttt{add\_resp} command. In addition to the 1st order RMF and ARF, we load the higher order responses to the HRC-LETG spectrum while doing the spectral analysis in \texttt{XSPEC}.  

\begin{figure}
\centering
\includegraphics[trim=0 0 0 0, clip, scale=0.525]{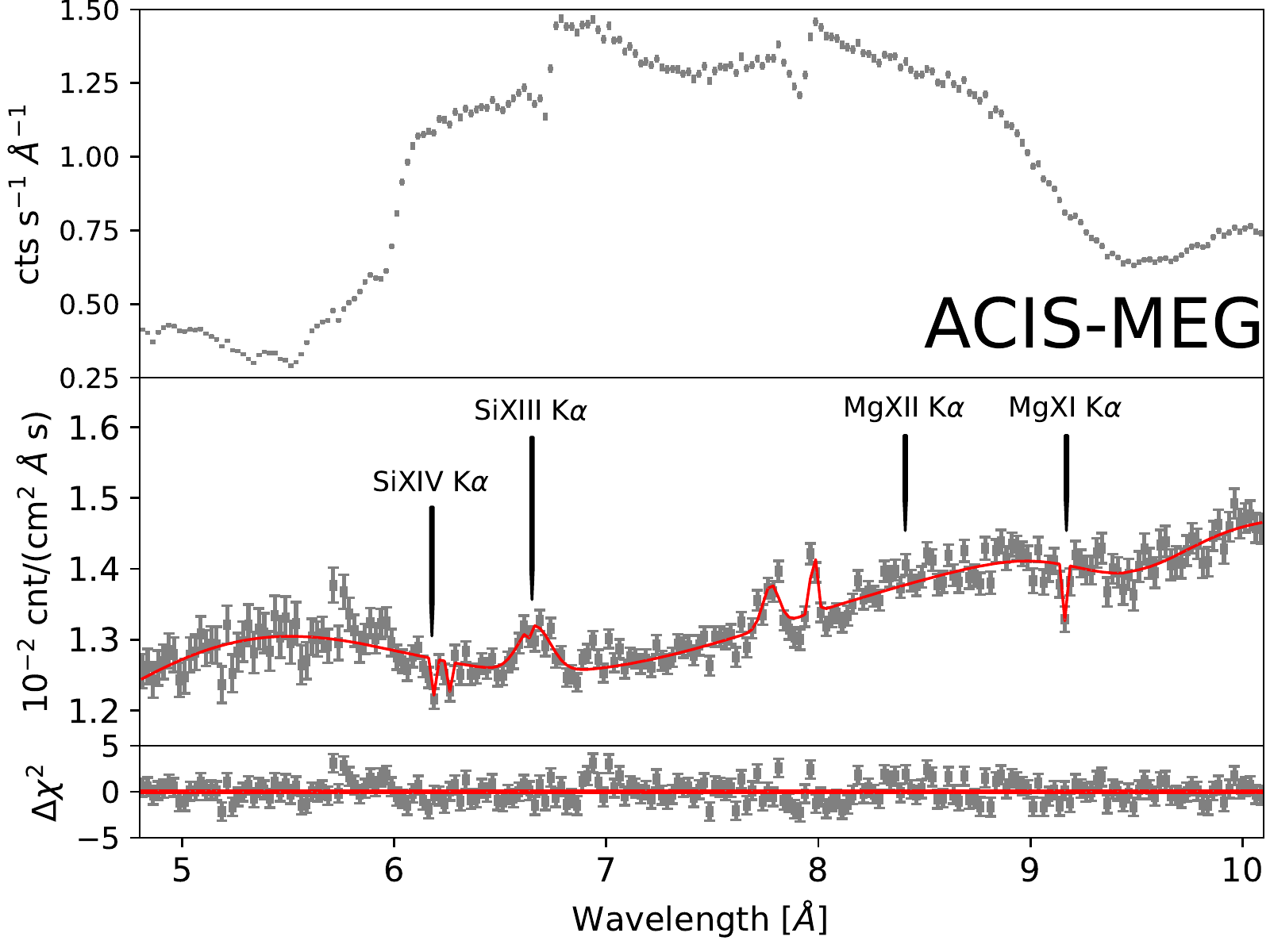}
\includegraphics[trim=0 0 0 0, clip, scale=0.525]{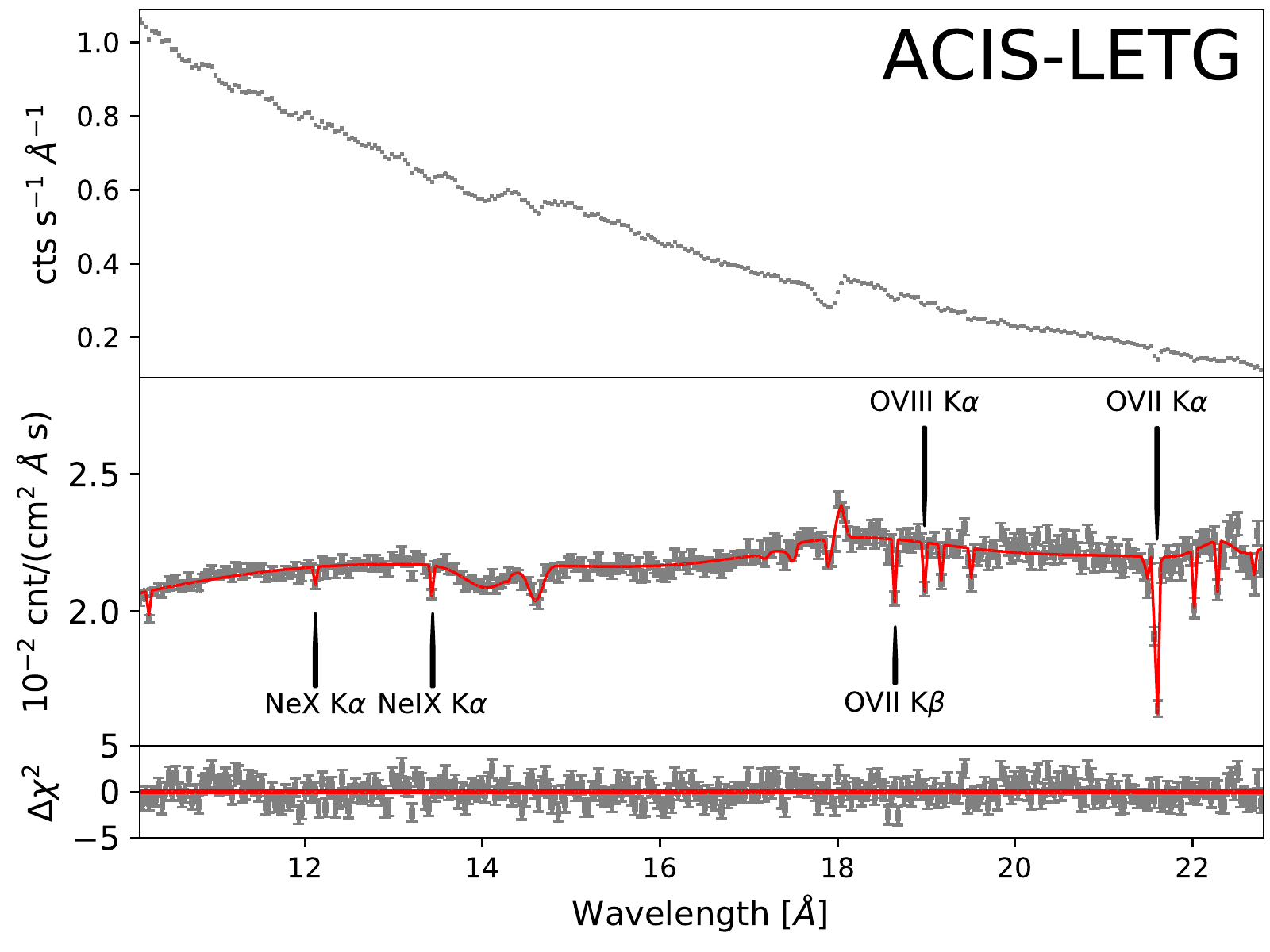}
\includegraphics[trim=0 0 0 0, clip, scale=0.525]{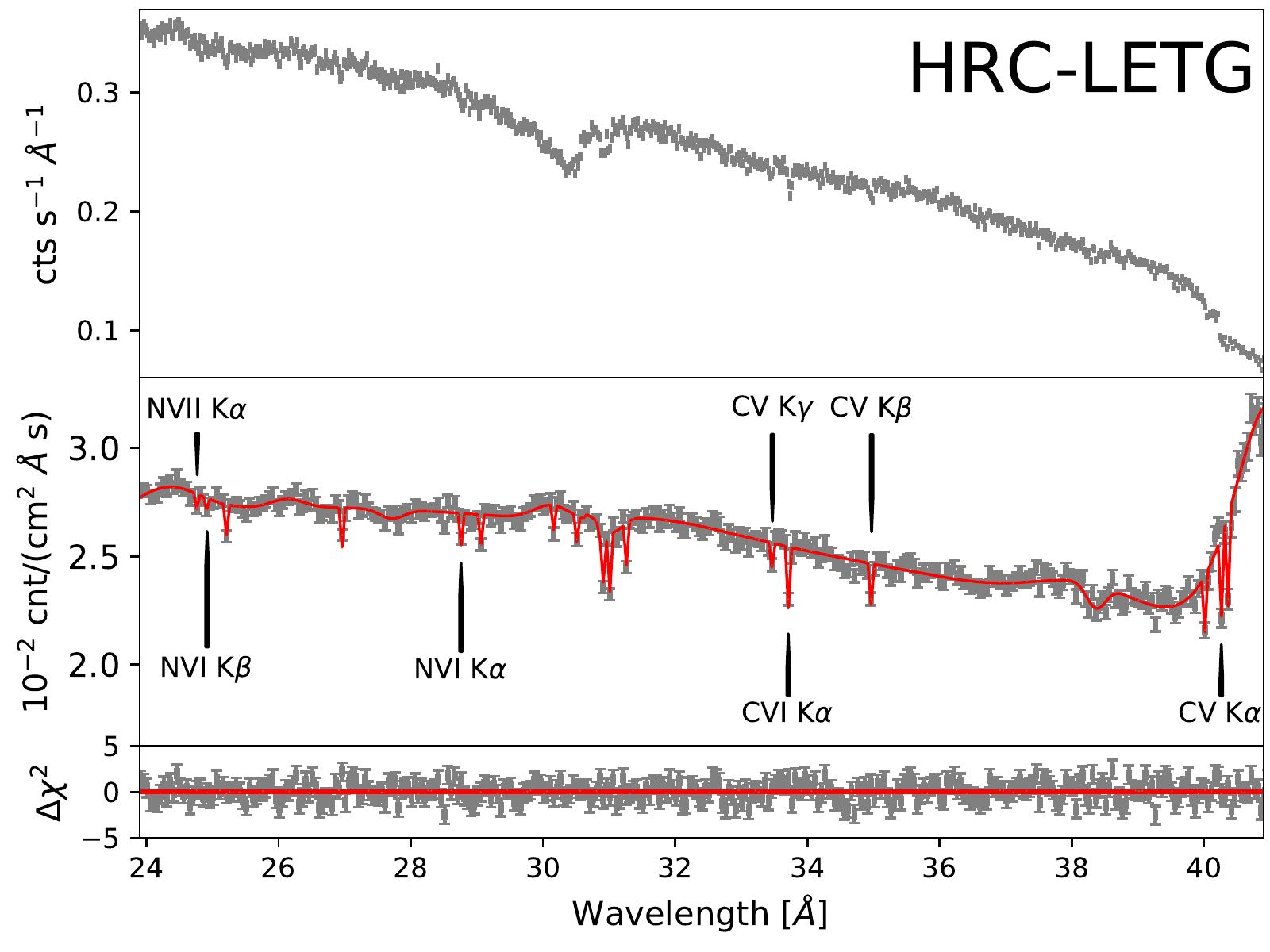}
\caption{\label{fig:spectrum} Each subplot: the data, {i.e., the spectra folded with the effective area and LSF} (top), and unfolded spectra (middle) of Mrk\,421, in bins with a S/N per bin $\geqslant 150$. The best-fit continuum (\S\ref{sec:identify}) with $z=0$ absorbers modeled with Gaussian lines are shown. The residuals are plotted in the bottom panels. The location of the \sixiii absorption line is on top of a broad Gaussian continuum feature (see \S\ref{sec:identify}); it is not identified as a $z=0$ emission feature. }
\end{figure}

The total exposure time of ACIS-MEG, ACIS-LETG and HRC-LETG data are 283.5\,ks, 627.8\,ks and 604.7\,ks, respectively. This yields X-ray grating spectra with signal-to-noise ratio per resolution element (SNRE)\footnote{the resolution element is defined as half-width half maxima (HWHM) of the line spread function (LSF), which is 12.5 m\AA~and 25 m\AA~for MEG and LETG, respectively.} of 59, 87 and 62, presented in figure \ref{fig:spectrum}. Strong lines of highly ionized metals at $z=0 $ are clearly seen at the expected wavelengths (Table \ref{tab:lineprofiles}). 

\subsection{$z$=0 lines: ionization model-independent results}\label{sec:identify}
\noindent The first step of detecting and characterizing the absorption lines is to determine the source continuum. The continuum of Mrk\,421 has varied remarkably over 20 years, both in shape and normalization. { Fitting a global continuum over a large range of wavelength is notoriously challenging, therefore} we fit the portions of the continuum spectra in ACIS-MEG (5--10 \AA), ACIS-LETG (10--24 \AA) and HRC-LETG (24--43 \AA) separately. {We discuss the continuum modeling in detail in appendix \ref{app:continuum}.} Our focus here is the absorption lines of He-like and H-like metal ions, not the details of the continuum shape; they do not affect our analysis. 

{After fitting the continuum and other line-like residuals}, we model the transitions of He-like and H-like ions of C, N, O, Ne, Mg and Si with unresolved absorbing Gaussian profiles (\texttt{agauss}). We allow the central wavelengths to vary within the resolution element around their theoretical $z=0$ values. 

We detect the absorption lines of \cv K$\alpha$ (6.4$\sigma$), K$\beta$ (4.2$\sigma$) and K$\gamma$ (2.7$\sigma$); \cvi K$\alpha$ (6.0$\sigma$); \nvi K$\alpha$ (3.0$\sigma$); \ovii K$\alpha$ (18.1$\sigma$) and K$\beta$ (6.5$\sigma$); \oviii K$\alpha$ (5.0$\sigma$); \neix K$\alpha$ (4.7$\sigma$); \nex K$\alpha$ (3.2$\sigma$); \mgxi K$\alpha$ (4.2$\sigma$) and \sixiv K$\alpha$ (3.6$\sigma$) (figure \ref{fig:spectrum}).  We also provide 3$\sigma$ upper limits on \nvi K$\beta$, \nvii K$\alpha$, \mgxii K$\alpha$ and \sixiii K$\alpha$\footnote{we define the single-line statistical significance as EW/$\Delta$(EW), where $\Delta$(EW) is the 1$\sigma$ error of EW in the -ve side.}. {We list the equivalent widths (EWs) in Table \ref{tab:lineprofiles}.}

\begin{table*}
\renewcommand{\thetable}{\arabic{table}}
\centering
\caption{Absorption line parameters with 1$\sigma$ errors. For lines with $<2\sigma$ significance, we quote 3$\sigma$ upper limits. The 8th column is the summation of the 5th--7th columns, which are the results of \texttt{PHASE} modeling. The equivalent widths and corresponding column densities obtained from Gaussian line fitting are in the 4th and the 9th column, respectively. The column densities of the $\alpha$ and $\beta$ transitions of \cvn, \nvi and \ovii from Gaussian line-fitting have been shown separately in the 9th column.} \label{tab:lineprofiles}
\begin{tabular}{c c c c c c c c c }
\toprule
 Ion & Transition & $\lambda$ &  EW & N$_{T_1}$ & N$_{T_2}$ & N$_{T_3}$ & N$_{tot}$ & N$_{EW}$ \\
 & & (\AA) & (m\AA) & $(cm^{-2})$ & $(cm^{-2})$ & $(cm^{-2})$ & $(\times 10^{15} cm^{-2})$ & $(\times 10^{15} cm^{-2})$ \\
 (1) & (2) & (3) & (4) &  (5) & (6) & (7) & (8) & (9)\\
\midrule
\cv &  &  & & 1.6$^{+0.2}_{-0.4}\times 10^{15}$  & 2.3$\pm0.2\times 10^{14}$ & \ldots  & 1.87$^{+0.24}_{-0.45}$ & \st{5.70}$_{-1.12}^{+2.27}\dagger$ \\
& K$\alpha$ & 40.27 & 7.16$\pm1.12$ &  &  &  & & 0.78$\pm0.12$\\ 
& K$\beta$ & 34.97 & 4.17$\pm0.79$ &  &  &  & & 2.74$\pm0.52$\\
& K$\gamma$ & 33.43 & 2.07$\pm0.78$ &  &  &  & & 3.93$\pm1.48$ \\
\cvi & K$\alpha$ & 33.74 & 5.98$\pm0.78$ & 1.0$^{+0.2}_{-0.3}\times 10^{12}$ & 1.3$\pm0.1\times 10^{15}$ & 2.0$\pm0.6\times 10^{14}$ & 1.50$\pm0.14$ & 1.43$\pm0.18$\\
\nvi &  &  &  & 1.3$^{+0.2}_{-0.4}\times 10^{14}$  & 3.9$\pm0.4\times 10^{14}$ & 9.8$^{+3.0}_{-3.2}\times 10^{10}$ & 0.52$^{+0.04}_{-0.05}$ & \st{2.09}$^{+1.49}_{-0.57}\dagger$\\
& K$\alpha$ & 28.79 & 3.04$\pm0.68$ & & & & & 0.62$\pm0.14$ \\
& K$\beta$ & 24.90 & $<3.04$ & & & & & $<3.88$ \\
\nvii & K$\alpha$ & 24.78 & $<3.09$ & 2.2$^{+0.3}_{-0.6}\times 10^{10}$  & 4.6$\pm0.4\times 10^{14}$ & 1.0$\pm0.3\times 10^{14}$ & 0.56$\pm0.05$ & $<$1.37\\ 
\ovii &  &  & & 4.6$^{+0.7}_{-1.2}\times 10^{14}$ & 5.3$\pm0.5\times 10^{15}$ & 4.4$^{+1.3}_{-1.4}\times 10^{12}$ & 5.74$^{+0.52}_{-0.50}$ & \st{14.87}$^{+0.89}_{-0.70}\dagger$\\ 
& K$\alpha$ & 21.60 & 14.44$\pm0.81$ &   &  &  & & 5.02$\pm0.28$\\
& K$\beta$ & 18.63 & 3.82$\pm0.57$ &  &  &  & & 8.52$^{+1.30}_{-1.24}$\\ 
\oviii & K$\alpha$ & 18.97 & 2.96$\pm0.5$ & 2.3$^{+0.3}_{-0.6}\times 10^{10}$ & 9.3$^{+0.9}_{-0.8}\times 10^{14}$ & 1.6$\pm0.5\times 10^{15}$ & 2.51$_{-0.51}^{+0.49}$ & 2.24$\pm0.44$\\ 
\neix & K$\alpha$ & 13.45 & 1.94$^{+0.44}_{-0.40}$ & \ldots & 1.5$\pm0.1\times 10^{15}$ & 4.5$\pm1.4\times 10^{13}$ & 1.53$^{+0.15}_{-0.14}$ & 1.68$^{+0.38}_{-0.35}$\\ 
\nex & K$\alpha$ & 12.13 & 1.14$^{+0.38}_{-0.36}$ & \ldots & 2.8$\pm0.3\times 10^{12}$ & 3.0$^{+0.9}_{-1.0}\times 10^{15}$ & 2.98$^{+0.90}_{-0.96}$ & 2.12$^{+0.70}_{-0.66}$ \\ 
\mgxi & K$\alpha$ & 9.17 & 1.41$\pm$0.34 & \ldots & 3.4$\pm0.3\times 10^{15}$ & 4.2$^{+1.3}_{-1.4}\times 10^{11}$ & 3.42$^{+0.33}_{-0.31}$ & 2.57$\pm$0.61$\ddagger$\\
\mgxii & K$\alpha$ & 8.41 & $<$0.19 &  \ldots & 9.1$\pm0.9\times 10^{10}$ & 1.0$\pm0.3\times 10^{13}$ & 0.010$\pm$0.003 & $<$0.73\\
\sixiii & K$\alpha$ & 6.65 & $<$1.18 & \ldots & 3.7$^{+0.4}_{-0.3}\times 10^{13}$ & 8.5$^{+2.6}_{-2.7}\times 10^{14}$ & 0.89$^{+0.26}_{-0.27}$ & $<$4.02\\
\sixiv & K$\alpha$ & 6.18 & 1.05$\pm$0.23 & \ldots & \ldots & 8.5$^{+2.6}_{-2.7}\times 10^{15}$ & 8.53$^{+2.6}_{-2.7}$ & 7.48$\pm$2.08\\
\bottomrule
\end{tabular} 
$\dagger$ \footnotesize{The absorption by \cvn, \nvi and \ovii have multiple temperature components of different strength. Therefore, the column density obtained by combining K$\alpha$ and K$\beta$ transitions of the same ion are not applicable; these are shown as numbers with a line through.} 
\\
$\ddagger$ \footnotesize{{By simultaneously fitting ACIS-MEG, ACIS-LETG and HRC-LETG, the EW = $0.63\pm0.19\pm0.14$\,m\AA, N = $1.14\pm0.34\pm0.26 \times$ 10$^{15}$ cm$^{-2}$. By simultaneously fitting ACIS-MEG and HEG, the EW = $0.90\pm0.26\pm0.01$\,m\AA, N = $1.64\pm0.46\pm0.01 \times$ 10$^{15}$ cm$^{-2}$. The first error bars are average statistical uncertainties and the second error bars are  systematic uncertainties due to multiple gratings/instruments.}}
\end{table*} 

{While estimating the continuum, we took a conservative approach and did not mask the wavelengths around the He-like and H-like metal ions. To test if this biases our results, we refit the continuum by removing $\pm0.1$\AA~region around each ion. Then we add these wavelength regions back and re-calculate the equivalent widths. From carbon to silicon, the best-fitted EWs of the ions detected with $>2\sigma$ are 6.66, 4.13, 2.06, 5.80, 3.03, 14.65, 3.88, 3.03, 1.65, 1.15, 1.26, and 1.12 m\AA, respectively. These are consistent with the previously obtained  best-fitted values within 1$\sigma$ (see Table \ref{tab:lineprofiles}). The upper limits of \mgxii and \sixiii are similarly consistent as well. The evidence of \nvi K$\beta$ and \nvii K$\alpha$ lines become stronger as the detection significance increases to $2.0\sigma$ due to the new, larger best-fitted EWs (from 1.14 m\AA~to 1.29 m\AA, and from 1.10 m\AA~to 1.26 m\AA, respectively). The best-fitted EWs of all ions do not increase or decrease systematically due to masking the surrounding wavelengths. This shows that the estimation of EWs and the quantities derived from it are not affected by the approach of continuum fitting (see Appendix \ref{app:continuum}) for details.}

We calculate the column densities (or the upper limits for the undetected ones) of the ions from their respective equivalent widths assuming that they lie on the linear regime of the curve-of-growth (table \ref{tab:lineprofiles}). We also estimate the column densities of \cvn, \nvi and \ovii by combining the information of their K$\alpha$ and K$\beta$ lines, as has  been done previously by \cite{Williams2005,Gupta2012,Nicastro2016a,Gupta2017}. N(\cvn), N(\nvin) and N(\oviin) calculated this way are larger than those calculated in the curve-of-growth analysis (table \ref{tab:lineprofiles}), indicating that their K$\alpha$ lines are likely saturated. 

The absorption lines we detect, or obtain an upper limit on, suggest several interesting aspects of the observed system in terms of the temperature and the abundance ratios of metals, which we discuss below. 


The $b$-parameters of \cv and \nvi are similar: $b_C = 15\pm3$\,km s$^{-1}$, $b_N = 11\pm3$\,km s$^{-1}$, but they are different from the $b$-parameter of \ovii: $b_O = 73\pm5$\,km s$^{-1}$. Assuming the velocity broadening to be purely thermal, it corresponds to the temperature of T$_C$ = $1.7\pm0.6\times$10$^{5}$\,K, T$_N$ = $1.0\pm0.6\times$10$^{5}$\,K and T$_O$ = $5.1\pm0.7\times$10$^6$\,K. This suggests that the detected carbon, nitrogen and oxygen lines are not coming from the same phase; there are either multiple temperature components and/or non-thermal line broadening.

In collisional ionization equilibrium (CIE), the ionization fraction of He-like ions (e.g., \oviin) are plateaued at a high value ($\geqslant 0.8$) across a definite range of temperatures. Similarly, the ionization fraction of H-like ions (e.g., \oviiin) peak at definite temperatures. Therefore, the column density ratio of the same element's two ions can uniquely determine the temperature (T$_{ratio}$), assuming the ions are exclusively coming from a single phase. We estimate the temperature (or its limit) for each element individually.  

The \cv to \cvi ratio implies T $= 7.1\pm0.4\times10^5$\,K; \nvi to \nvii ratio implies T $= 1.1\pm0.2\times10^6$\,K; \ovii to \oviii ratio implies T $= 1.5\pm0.1\times10^6$\,K; \neix to \nex ratio implies T $= 3.1\pm0.4\times10^6$\,K; \mgxi to \mgxii ratio implies T $<7\times10^6$\,K, and \sixiii to \sixiv ratio implies T $= 2.8^{+1.0}_{-0.8}\times10^7$\,K. Clearly, the temperature windows do not overlap with each other. This suggests that a single, or even two-temperature model may be inadequate to characterize the observed spectra. 

Additionally, the column density ratio of two elements provide an approximate abundance ratio of those elements. This is valid under the assumption that both the elements are from the same phase and are co-spatial. We calculate the column density of an element using the measured column density of the ion with higher significance (e.g., \nvi for nitrogen and \sixiv for silicon), and the ionization fraction of that ion at T$_{ratio}$. For Mg, the upper limit of temperature is not constraining enough, so we assume N(Mg)$\geqslant$ N(\mgxin), leading to a lower limit on Mg/O. It is evident from the temperature estimates of C and Si that to measure [C/O] and [Si/O] we need to correct for fully ionized carbon and oxygen, requiring ionization modeling.  Therefore, we cannot estimate [C/O] and [Si/O] in a model-independent way. 

Given the solar abundance ratios from \cite{Asplund2009}, we find that [N/O] = 1.1$\pm$0.3\,N/O$_\odot$, [Ne/O] = 1.5$^{+0.8}_{-0.6}$\,Ne/O$_\odot$ and Mg/O$\geqslant$ 1.8$^{+0.7}_{-0.6}$\,Mg/O$_\odot$. This suggests that nitrogen is in solar mixture with oxygen, but neon and magnesium are mildly super-solar relative to oxygen in the observed phase. The \fexviin--\fexxv lines and the Fe unresolved transition array (UTAs) of \fexvin--\fexviiin~ are not detected at better than $1\sigma$ significance. This suggests $\alpha$-enhancement relative to iron. 

\subsection{\texttt{PHASE} modeling of the $z=0$ absorbers}\label{sec:phase}
\noindent To confirm and to quantitatively determine the suggestive results obtained in \S\ref{sec:identify}, we now model the data in detail. We use the hybrid-ionization model \texttt{PHASE} \citep[models of collisionally-ionized gas perturbed by photo-ionization by the meta-galactic radiation field, at a given redshift;][]{Krongold2003,Nicastro2018,Das2019a} to fit the data. The free parameters of the \texttt{PHASE} model are the temperature T, equivalent hydrogen column density N(H), relative abundance of the metals in the absorbers, non-thermal line broadening $b_{\rm nT}$, photo-ionization parameter $U$ and redshift $z$ (equivalent to line-of-sight velocity). The model assumes relative abundances and absolute metallicities to be solar by default, but allows them to vary between 0.01 to 100 times solar, independently for each elements: He, C, N, O, Ne, Mg, Al, Si, S, Ar, Ca, Fe and Ni. 

In our models, we allow T, N(H), $b_{\rm nT}$ and relative abundance of the metals to vary. C, N, O, Ne, Mg and Si have prominent absorption features (figure \ref{fig:spectrum}), and Fe lines e.g., \fexix (13.425\AA), \fexvii (12.123\AA), \fexxi (12.165\AA), \fexxii~(9.163\AA~and 9.183\AA)~can contaminate the \neixn, \nex and \mgxi lines through blending. Therefore, we vary the abundances of C, N, Ne, Mg, Si and Fe with respect to O. The spectrum is insensitive to the abundances of other elements, so we freeze their abundances with respect to oxygen at solar. The CGM is assumed to be in CIE, therefore we freeze $U$ to the lowest possible value allowed by \texttt{PHASE}, at $U$=$10^{-4}$, ensuring that photo-ionization is negligible\footnote{\textit{U} is the flux of ionizing photons per unit density of gas: $\textit{U} = \frac{Q(H)}{4\pi r^2n(H)c}$, where Q(H) is the number of hydrogen ionizing photons s$^{-1}$, r is the distance to the source, n(H) is the number density of hydrogen.}. The best-fit central wavelengths of the absorbers are consistent with their theoretical values within spectral resolution, indicating no significant line-of-sight velocity. Therefore, we keep $z$ frozen at 0. 

To begin with, we fit the whole spectrum\footnote{{ACIS-MEG (5--10 \AA), ACIS-LETG (10--24 \AA) and HRC-LETG (24--43 \AA) simultaneously}} with a single-temperature \texttt{PHASE} model with solar composition. The best-fit model ($\chi^2/dof$ = 3757.93/3625) is a significant improvement over the continuum ($\chi^2/dof$ = 4275.16/3628), as is expected from the existence of the absorption lines. This model severely underestimates the \mgxi line although it is expected to be present at the best-fit temperature. This indicates that either the abundance ratios are non-solar and/or another temperature component is needed. Therefore, we fit the spectra with a two-temperature model, and solar composition. This model provides a better fit ($\chi^2/dof$ = 3747.64/3622) than the single-temperature solar composition model, showing that multiple temperatures are necessary. Next, we fit the spectra with a single temperature model, but with non-solar composition (\texttt{PHASE\_A}). This also improves the fit ($\chi^2/dof$ = 3735.28/3620), but it cannot reproduce all the detected lines in the spectra, in particular it does not account for the \cvn, \nex and \sixiv lines. 

To see if the non-solar abundance ratios are necessary in the multiple temperature scenario, we fit the spectra with two \texttt{PHASE} models (\texttt{PHASE\_A*PHASE\_B}). We allow the H column density and the temperatures in two phases to be different, but do not force them to be different. Most of the metal ions of interest probe only one of the two temperatures (e.g., carbon and nitrogen probe the lower temperature while silicon probes the higher temperature), therefore we force the relative abundances of light elements ([C/O], [N/O] and [Si/O]) to be the same in both phases. Ne, Mg and Fe, on the other hand, have lines in both the temperature components (see \S\ref{sec:identify}), so we allow [Ne/O], [Mg/O] and [O/Fe] to be different in the two phases. The two-temperature non-solar composition model provides a better fit ($\chi^2/dof$ = 3707.30/3614) than the two-temperature solar composition model, and it accounts for the \nex and \sixiv lines. 

The two-temperature non-solar composition model cannot reproduce the \cv lines, and it underestimates the \nvi line. This is not unexpected; the b-parameters of \cv and \nvi indicated a lower temperature than that indicated by all other ions (see \S\ref{sec:identify}). Therefore, we further add another \texttt{PHASE} component (\texttt{PHASE\_C}). We do not force the temperature and the H column density of this phase to be different from the other two phases. We allow [C/O] to be different from the other phases because \cv can independently constrain this temperature. There are no detectable Ne, Mg and Fe lines  in the expected temperature range of the third component, so we assume the [Ne/O], [Mg/O] and [O/Fe] ratios to be solar in this phase. The nitrogen lines are too weak to independently constrain the two temperatures (one dominated by \nvii and the other by \nvin), so we tie [N/O] of this phase to be the same as the first phase.

The three-temperature non-solar composition model (\texttt{PHASE\_A*PHASE\_B*PHASE\_C}) fits the data better ($\chi^2/dof =3673.58/3610$) than the previous models, reproduces all the detected lines, and is consistent with the non-detections. This is our final spectral fit. The best-fitted parameter values in \texttt{PHASE} models are quoted in table \ref{tab:best-fit}. 

\begin{table}
\renewcommand{\thetable}{\arabic{table}}
\centering
\caption{Parameters of the best-fitted \texttt{PHASE} model (\texttt{PHASE\_C}, \texttt{PHASE\_A} and \texttt{PHASE\_B} are mentioned in the increasing order of their best-fitted temperatures) with 90\% confidence interval and 3$\sigma$ upper/lower limit. The column densities are quoted for solar metallicity of oxygen. Abundances are {in log$_{10}$} with respect to the solar composition, according to the prescription of \cite{Asplund2009}}
\label{tab:best-fit}
\begin{tabular}{c c}
\toprule
Parameters & Values \\
\midrule 
T$_1$ & $3.0\pm0.4\times 10^5$\,K
\\ 
T$_2$  & $1.5\pm 0.1\times 10^6$\,K
\\
T$_3$  & $3.2^{+1.5}_{-0.5}\times 10^7$\,K
\\
\tableline
N(H)$_1$ & 1.1$^{+0.2}_{-0.3}$ $\times 10^{18}$\,cm$^{-2}$ \\
N(H)$_2$ & 8.4$\pm$0.8 $\times 10^{18}$\,cm$^{-2}$ \\
N(H)$_3$ & $9.1^{+2.7}_{-2.9} \times 10^{20}$\,cm$^{-2}$ \\
\tableline
$b_{nT,1}$ & $<31$\,km s$^{-1}$   \\
$b_{nT,2}$ & 221$^{+106}_{-86}$\,km s$^{-1}$    \\
$b_{nT,3}$ & 711$^{+594}_{-411}$\,km s$^{-1}$    \\
\tableline
\multicolumn{2}{c}{Non-solar}\\
\tableline
{
{$[$C/O$]_1$}} & {{0.60}$^{+0.14}_{-0.10}$}\\
{{$[$C/O$]_2$}} & {{0.24}$^{+0.12}_{-0.10}$}  \\
{{$[$Mg/O$]_2$}} & {{1.11}$^{+0.12}_{-0.17}$}$\dagger$   \\
{{$[$Mg/Fe$]_2$}} & {{1.06}$^{+0.12}_{-0.34}$}$\dagger$   \\
{{$[$Ne/O$]_{3}$}} & {{0.42}$\pm{0.20}$}  \\
{{$[$O/Fe$]_3$}} & {$>${0.66}}  \\
{{$[$Ne/Fe$]_{3}$}} & {$>${0.82}}  \\
{{$[$Si/Fe$]_3$}} & {$>${0.68}}  \\
\tableline
\multicolumn{2}{c}{Solar}\\
\tableline
{$[$N/O$]_{1/2}$} & {$0.10^{+0.17}_{-0.15}$}  \\
{$[$Ne/O$]_{2}$} & {0.19$\pm$0.18}  \\
{$[$O/Fe$]_2$} &  {$-0.06^{+0.43}_{-0.28}$}  \\
{$[$Si/O$]_3$} & {0.34$\pm$0.23}  \\ 
\bottomrule
\end{tabular} 

$\dagger$ \footnotesize{{By considering the EW estimated from simultaneous fitting of ACIS-MEG, ACIS-LETG and HRC-LETG, the [Mg/O] and [Mg/Fe] are 0.76 and 0.71, respectively. Similarly, by considering the EW estimated from simultaneous fitting of ACIS-MEG and HEG, the [Mg/O] and [Mg/Fe] are 0.92 and 0.86, respectively.}}
\end{table} 
We have allowed the non-thermal broadening of metal lines $b_{\rm nT}$ to vary throughout our analysis. To test if the non-thermal components are really necessary, we refit the spectrum with $b_{\rm nT}$ in all phases forced to be zero, and find that the best-fit model becomes worse: $\chi^2/dof =3698.50/3613$. This  confirms that the non-thermal broadening is required to fit the spectrum\footnote{$\Delta \chi^2$ after forcing $b_{\rm nT}$ to be zero in \texttt{PHASE\_A}, \texttt{PHASE\_B}, and \texttt{PHASE\_C} individually are 20.99, 5.41 and 0.77, respectively. This shows the relative importance of non-thermal broadening in each phase.}. 

\begin{figure}
    \centering
    \includegraphics[scale=0.53]{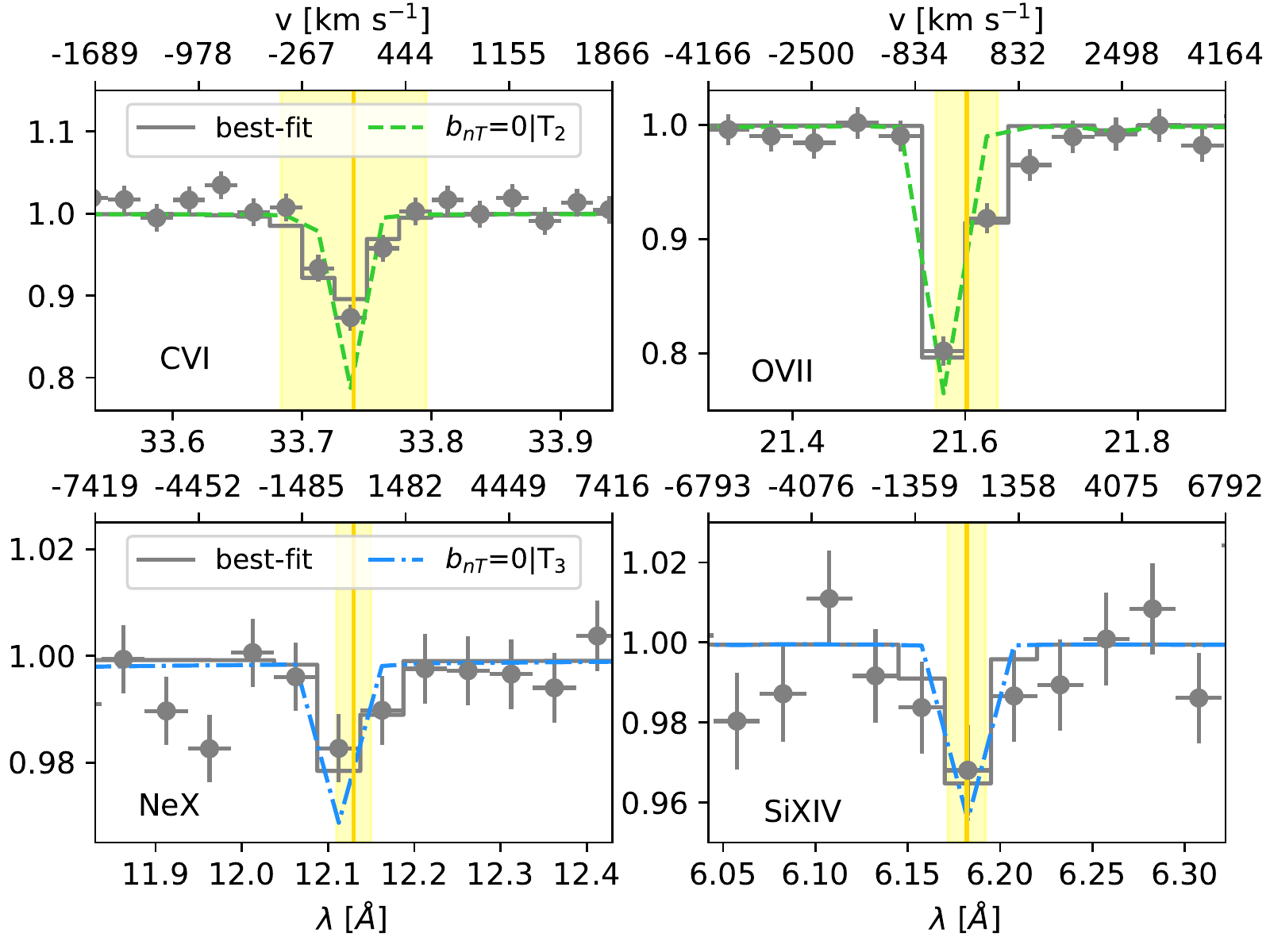}
    \caption{Non-thermal line broadening in the \textit{warm-hot} (T$_2$) and the \textit{hot} (T$_3$) phases. The lines are unresolved, making the effect of non-thermal broadening not prominent visually, even though it is required by the spectral fits. On the top axes of each panel we show the line-of-sight velocities in the Galactic standard of rest (GSR). The vertical lines and the shaded regions correspond to  v$_{\rm GSR}=0$ and v$_{\rm GSR}=\pm500$ km s$^{-1}$ respectively.}
    \label{fig:non-thermal}
\end{figure}

The spectral resolution corresponds to a velocity resolution $\Delta$v of $\approx$250 km\,s$^{-1}$ around \nvi and \cv lines (tracers of \texttt{PHASE\_C} at T$_1$), $\Delta$v of 220, 350, 560 and 410 km\,s$^{-1}$ around \cvin, \oviin, \neix and \mgxi lines (tracers of \texttt{PHASE\_A} at T$_2$) respectively, and $\approx$610 km\,s$^{-1}$ around \sixiv and \nex lines (tracers of \texttt{PHASE\_B} at T$_3$). The thermal broadening $b_{\rm T}$ calculated from the best-fit temperatures (table \ref{tab:best-fit}) are 18--20 km\,s$^{-1}$, 32--46 km\,s$^{-1}$ and 128--221 km\,s$^{-1}$ for T$_1$, T$_2$ and T$_3$ respectively. By comparing the total broadening of the lines $\sqrt{(b_{\rm T})^2 + (b_{\rm nT})^2}$ in each phase with $\Delta$v, we can see that most of the ion lines are not resolvable (figure \ref{fig:non-thermal}). \nex and \sixiv should be partially resolvable at the best-fit value of $b_{\rm nT}$, but the large error on $b_{\rm nT}$ pushes it to the limit of not being resolvable. This is consistent with our model-independent results where we assume the lines to be unresolved (\S \ref{sec:identify}). 

\begin{figure}[h!]
\centering
\includegraphics[scale=0.5]{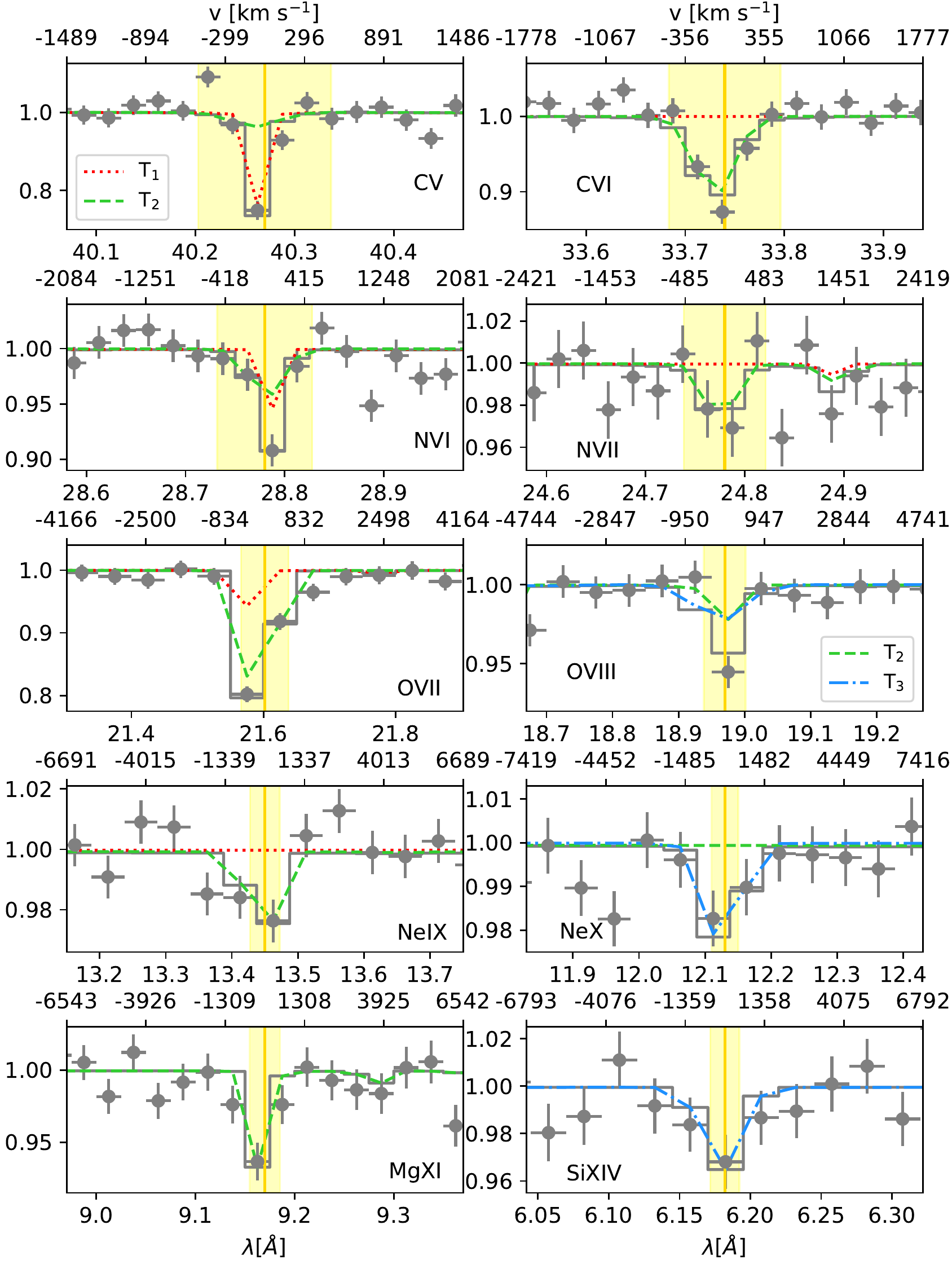}
\caption{\label{fig:components}The absorption (solid gray lines) decomposed into \textit{warm} (T$_1$-dotted red lines), \textit{warm-hot}   (T$_2$-dashed green lines) and \textit{hot}  (T$_3$-dash-dotted blue lines) phases shown separately for each of the detected metal transitions, normalized by the best-fitted continuum model. The top axes,  the vertical lines and the shaded regions are as in Figure \ref{fig:non-thermal}.}
\end{figure}
The non-thermal broadening of a line may be the manifestation of more than one thermally broadened lines, with the central wavelengths shifted due to non-zero line-of-sight velocity. The velocity resolution of the spectrum does not allow us to distinguish between the two cases of non-thermal broadening vs. multiple velocity components. This is consistent with the model-independent results (\S \ref{sec:identify}) where the observed wavelength of the lines did not indicate any line-of-sight velocity larger than the velocity resolution. 

The column densities of the detected ions from the three-temperature \texttt{PHASE} model are presented in table \ref{tab:lineprofiles} (5th--7th column). The sum of these column densities is given in the 8th column. Comparing these with the column densities estimated from the curve-of-growth analysis (9th column), we find that the two are consistent with each other within 2$\sigma$.    

We do not use the results of Gaussian line-fitting as a prior in the \texttt{PHASE} modeling. Therefore, fitting the data using \texttt{PHASE} is completely independent from the curve-of-growth analysis in the previous section. \texttt{PHASE} takes into account line saturation by Voigt profile fitting; the agreement between the two shows that the effect of saturation in the absorption lines, if any, is negligible for most of the ions. The K$\alpha$ line of \cv indicates saturation, because its column density from the curve-of-growth analysis is smaller than that from \texttt{PHASE}. However, the column densities obtained by combining K$\alpha$ and K$\beta$ lines of \cvn, \nvi and \ovii are larger than their respective column densities estimated from \texttt{PHASE}. The method of combining K$\alpha$ and K$\beta$ transitions for saturated lines is applicable only when the relevant absorption has one component \citep{Draine2011}. Given that \cvn, \nvi and \ovii contribute in multiple temperature components (figure \ref{fig:components}), the column densities estimated by considering K$\alpha$ and K$\beta$ lines together are not valid. This shows the power of \texttt{PHASE} modeling, where the physical parameters of the observed system can be estimated {without the confusion of single vs. multiple components}. 


\section{Results}\label{sec:results}
\begin{figure*}
    \centering
    \includegraphics[trim=5 0 0 0 ,clip,scale=0.8]{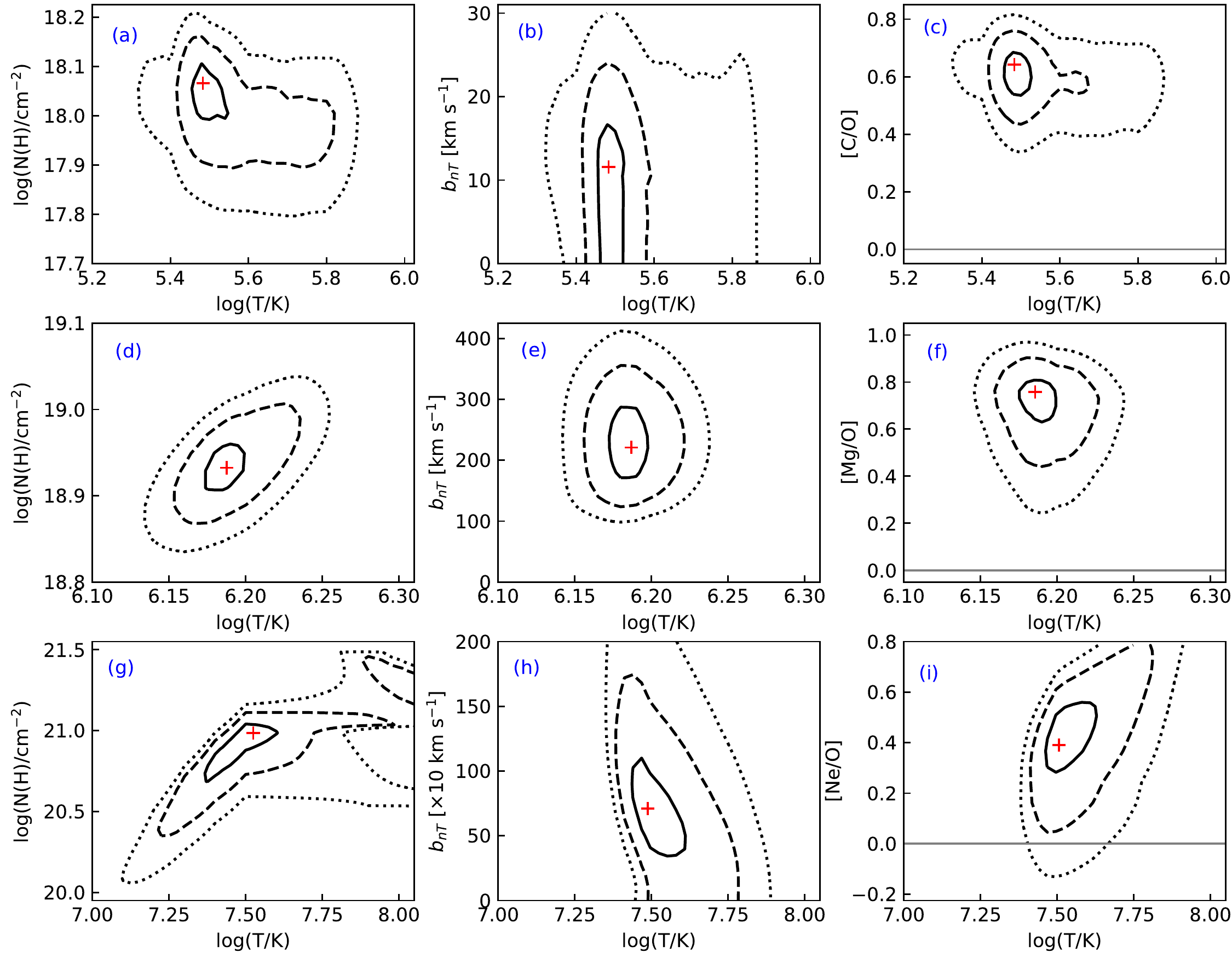}
    \caption{Contour plots of the three \texttt{PHASE} components. Top: \textit{warm}, middle: \textit{warm-hot}, bottom: \textit{hot} phase. Left: hydrogen column density (modulo solar metallicity of oxygen), middle: non-thermal broadening and right: abundance ratios with respect to oxygen, vs. the temperature of the respective phases. The `+' sign is the best-fit value and solid, dashed and dotted lines correspond to 68\%, 90\% and 99\% confidence intervals. The horizontal line in the right panels correspond to solar composition.}
    \label{fig:contour}
\end{figure*}
\noindent The best-fit \texttt{PHASE} model (table \ref{tab:best-fit}) has yielded 5 interesting results: 1) we have detected a \textit{hot}  gas phase probed by \sixiv and \nex absorption lines; 2) we have detected the \textit{warm-hot} phase probed by \cvin, \oviin, \oviiin, \neix and \mgxi absorption lines; 3) we have detected a \textit{warm} phase probed by \cv and \nvi absorption lines; 4) we have found super-solar $\alpha$/Fe in the \textit{hot} and the \textit{warm-hot} phases and non-solar abundance ratios of the light elements in all phases; and 5) we have constrained the non-thermal broadening in the \textit{hot} and the \textit{warm-hot} phases. Below, we discuss each result in details. 

\subsection{Temperature} 
The temperatures of the three phases differ from each other by more than 3$\sigma$ (figure \ref{fig:contour}), and span a temperature range of two orders of magnitude (table \ref{tab:best-fit}). \cv is present predominantly at the lower temperature (T$_1$ - \textit{warm} phase); \cvin, \nviin, \neix and \mgxi are present only at the intermediate temperature (T$_2$ - \textit{warm-hot} phase); \nex and \sixiv are present only at the higher temperature (T$_3$ - \textit{hot} phase). \nvi contributes to both the \textit{warm} and the \textit{warm-hot} phases, and \oviii contributes to both the \textit{warm-hot} and the \textit{hot} phases (figure \ref{fig:components}). \ovii predominantly comes from the \textit{warm-hot} phase, and it contributes significantly to the \textit{warm} phase, in which  our \texttt{PHASE} model predicts \ov and \ovi to dominate\footnote{N(\ovn) = 1.5$^{+0.2}_{-0.4}\times10^{14}$ cm$^{-2}$, N(\ovin) = 1.6$^{+0.2}_{-0.4}\times10^{14}$  cm$^{-2}$ and N(\oviin) = 4.6$^{+0.7}_{-1.2}\times10^{13}$ cm$^{-2}$ in the \textit{warm} phase}. 

\subsection{Column density} 
The equivalent hydrogen column density (modulo solar metallicity of oxygen) N(H) of the three phases are starkly different from one other (table \ref{tab:best-fit}). The \textit{warm} phase has the lowest N(H) and the \textit{hot} phase has the highest N(H). There is no noticeable correlation between the temperature and the N(H) in the \textit{warm} phase {(reflected by the contour shape in figure \ref{fig:contour}, panel a)}. The \textit{warm} phase is probed by \cvn, \nvi and \ovii (figure \ref{fig:components}), and the ionization fraction of these ions are almost constant in some portion of the temperature range concerned. That makes the N(H) estimate independent of the temperature in the \textit{warm} phase. The \textit{warm-hot} and the \textit{hot} phases are predominantly probed by \ovii and \sixiv (and, \nexn) respectively whose ionization fractions decrease with temperature in the ranges of consideration. Therefore, for a given column density of \oviin, \nex and \sixiv the N(H) increases with temperature in the \textit{warm-hot} and the \textit{hot} phases. As a result, temperature and N(H) are positively correlated in the \textit{warm-hot} and partially in the \textit{hot}  phase (figure \ref{fig:contour}, {panels d and g}). The N(H) is practically insensitive of the temperature in the \textit{hot}  phase above a temperature where all the tracer elements are almost completely ionized.

\subsection{Line broadening} 
 We detect non-thermal line broadening in the \textit{warm-hot} and \textit{hot} phases (table \ref{tab:best-fit}; figure \ref{fig:non-thermal}, middle column). The \textit{warm-hot} phase has a smaller non-thermal broadening than that of the \textit{hot} phase. The upper limit of the non-thermal broadening in the \textit{warm} phase is smaller than those in the other two phases. In the \textit{warm} and the \textit{warm-hot} phases, the b-parameters are constrained from multiple transitions of the same ion (e.g., the K$\alpha$ and K$\beta$ lines of \cv and \oviin). The temperature and the thermal broadening are constrained from the ratio of two ions of the same element (e.g., \oviii and \oviin). Thus, the excess in the b-parameter compared to the thermal component can be constrained independent of the temperature. As a result, the temperature and the non-thermal broadening are not correlated in the \textit{warm} and the \textit{warm-hot} phases (figure \ref{fig:contour}, panels b and e). On the other hand, the \textit{hot} phase is predominantly probed by one ion of a given element (e.g., \sixiv or \nexn). Therefore, for a given b-parameter, thermal and non-thermal broadening are anti-correlated with each other (figure \ref{fig:contour}, panel h). The ratios of the non-thermal and the thermal broadening are $6^{+3}_{-2}$, and 4$^{+3}_{-2}$ (90\% confidence interval) in the \textit{warm-hot} and \textit{hot} phases, respectively. 

\subsection{Abundance ratios} 
The observed spectral wavelength range does not contain any line or edge of hydrogen. Therefore, we cannot calculate the metallicity, i.e., the absolute abundances of metals with respect to hydrogen. However, we can determine the metal abundances with respect to each other. Oxygen is the most abundant metal in the solar composition, so we report abundance ratios of metals with respect to oxygen. We find that carbon is super-solar in the \textit{warm} and \textit{warm-hot} phases, magnesium is super-solar in the \textit{warm-hot} phase, and neon is super-solar in the \textit{hot} phase (table \ref{tab:best-fit}, figure \ref{fig:nonsolarfe}). {Other} elements are consistent with a solar mixture. 

{Because [Mg/O] is larger than [C/O] and [Ne/O] by more than a factor of 3, and Mg abundance depends solely on the \mgxi line, we investigate if the \mgxi line in ACIS-MEG is overestimated. The EW of the \mgxi line by simultaneously fitting the ACIS-HEG and ACIS-MEG spectra is consistent with the EW from only the ACIS-MEG spectrum within 2$\sigma$ (see the footnote of Table \ref{tab:lineprofiles}). However, the EW by simultaneously fitting the HRC-LETG, ACIS-LETG and ACIS-MEG spectra\footnote{MEG has higher spectral resolution and the LETG data have higher S/N. Therefore, they are complimentary to each other in this case.} is smaller than the EW from only the ACIS-MEG spectrum by more than a factor of 2 (Table \ref{tab:lineprofiles}).  Out of these three measurements, we consider the smallest EW (and the column density) of \mgxi as the final value as a conservative estimate. We adjust [Mg/O] in the table \ref{tab:best-fit} (see the footnote) and the figure \ref{fig:contour} (panel f) accordingly. Because the \textit{warm-hot} phase is dominated by oxygen, other physical properties of this phase remain unchanged. While it does not affect our result and the following discussion qualitatively, the revised [Mg/O] is at par with other super-solar abundance ratios.    }

[C/O] and [Mg/O] are not correlated and [Ne/O] is positively correlated with temperature (figure \ref{fig:contour}). 
The ionization fraction of \cv and \mgxi are $\sim$independent of temperature and the ionization fraction of \nex decreases with temperature in the ranges of the temperature concerned. Therefore, for a given column density of these ions, the behavior of the abundance ratios with temperature are not unexpected. 

\fexvin--\fexviiin~UTA and \fexxii-\fexxv lines are expected to be present at T$_2$ and T$_3$ respectively, but are not detected at a significance of better than $1\sigma$ (figure \ref{fig:nonsolarfe}, right panel). 
By using the upper limit of iron abundance, we can determine the lower limit of the  abundance ratios of detected elements relative to iron (table \ref{tab:best-fit}). In the \textit{warm-hot} phase, [Mg/Fe] is enhanced and in the \textit{hot} phase, [O/Fe], [Ne/Fe] and [Si/Fe] are enhanced. 

\section{Discussion}\label{sec:discussion}
\noindent The \textit{hot} phase together with the \textit{warm-hot} phase was first discovered unambiguously along the sightline toward 1ES\,1553+113 \citep{Das2019a}. In this paper we report the discovery of three temperature components: \textit{hot}, \textit{warm-hot} and \textit{warm}, for the first time; these have not been observed earlier either in emission or in absorption. Below, we discuss each phase, and its relation to the other two phases, in detail. We constrain the non-thermal line broadening of highly ionized CGM phases for the first time. We discuss the role of non-thermal sources in the context of the line broadening. The non-solar abundance ratios have also not been measured earlier in absorption except along the sightline toward 1ES\,1553+113 \citep{Das2019a}. We qualitatively interpret the chemical composition in terms of metal enrichment, mixing and depletion. These provide interesting insights on the Galactic thermal and chemical evolution, and may affect the mass estimation of metals and baryons in the Galactic halo.

\begin{figure*}
\begin{subfigure}
\centering
\includegraphics[scale=0.55]{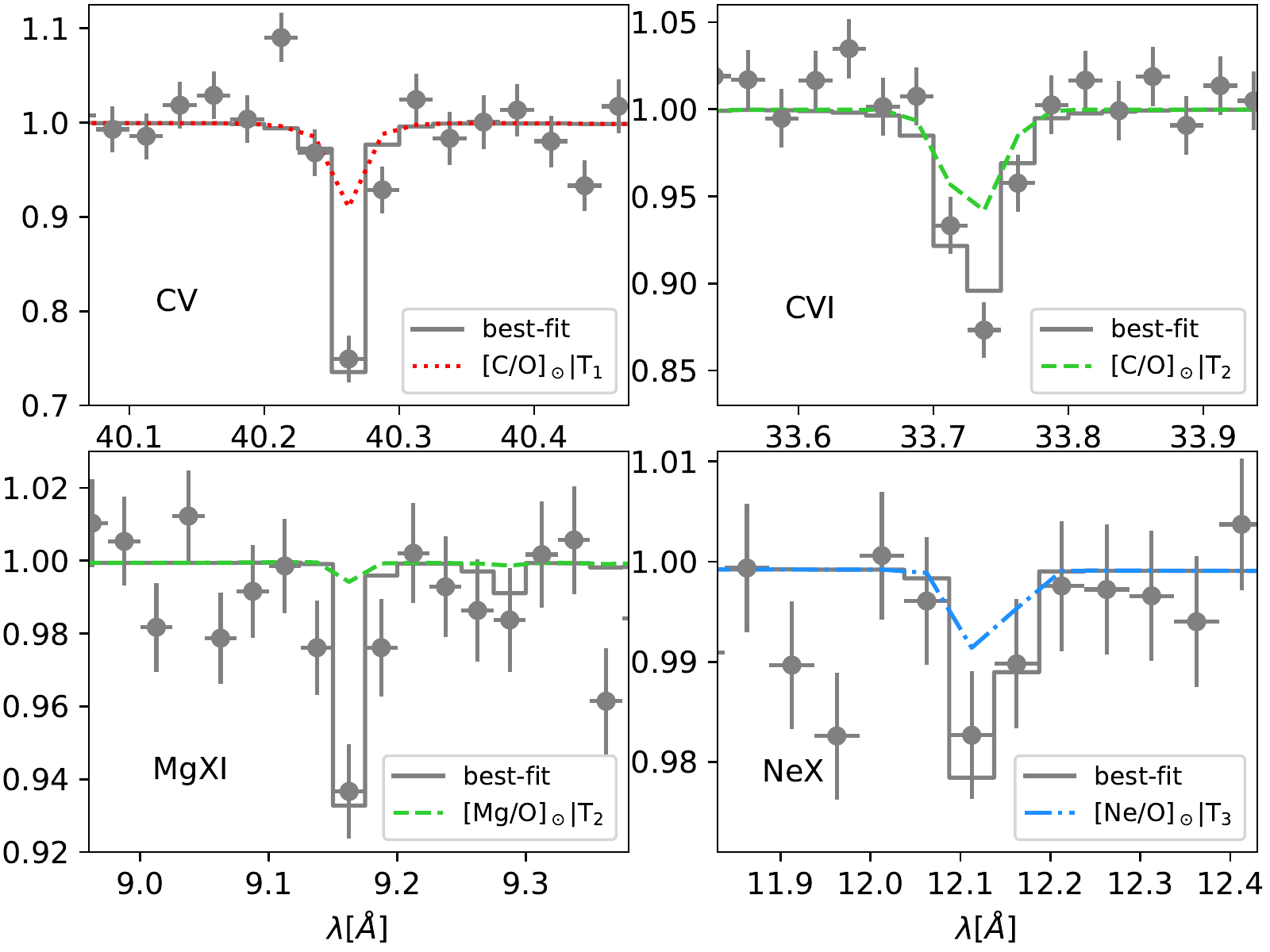}
\end{subfigure}
\begin{subfigure}
\centering
\includegraphics[scale=0.55]{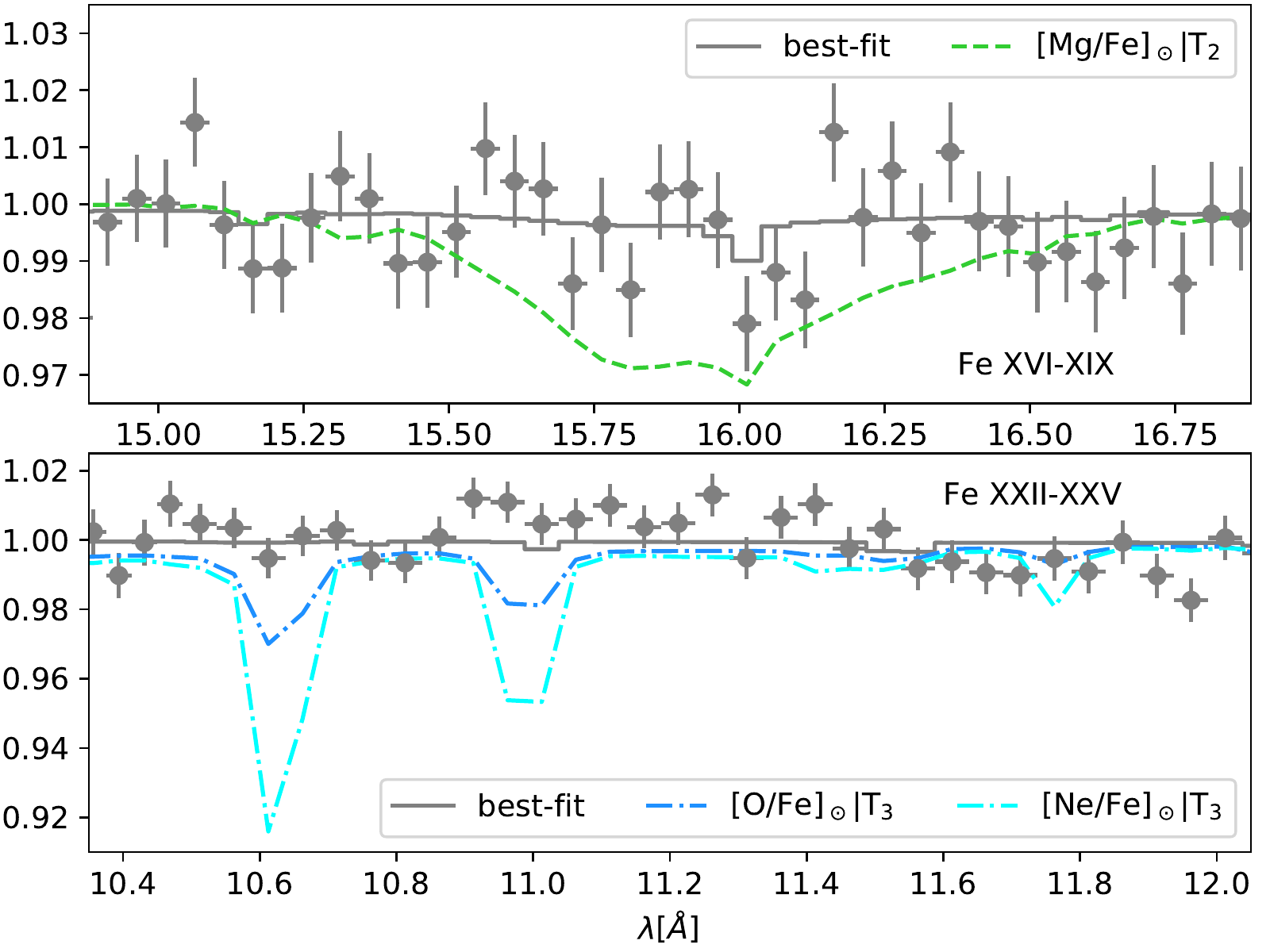}
\end{subfigure}
\vspace*{-0.5cm}
\caption{\label{fig:nonsolar}Non-solar abundance ratios in the \textit{warm}, \textit{warm-hot}  and \textit{hot}  CGM. Left: The detected absorption lines of carbon, neon, magnesium and silicon, normalized by the best-fitted continuum model, for the best-fit non-solar (gray solid lines) vs. solar (\textit{warm} in red dotted, \textit{warm-hot} in green dashed and \textit{hot}  in blue dash-dotted lines) abundance ratios with respect to oxygen. The observed spectra clearly require non-solar abundance ratios of $\alpha$-elements. \label{fig:nonsolarfe} Right: The absorption profile of Fe for best-fit non-solar vs. solar [Mg/Fe] in the \textit{warm-hot}   phase (green dashed lines), solar [O/Fe] (blue dash-dotted lines), solar [Ne/Fe] or [Si/Fe] (cyan dash-dotted lines) in the \textit{hot} phase. In the top panel, the broad feature corresponds to the Fe UTA that would have been produced in the \textit{warm-hot} phase with solar abundances of [Mg/Fe]. In the bottom panel, the lines correspond to \fexxii--\fexxv transitions produced in the \textit{hot}  phase, that would have been detected at the shown strength if [O/Fe] and [Ne/Fe] were solar. Super-solar $\alpha$/Fe in the \textit{hot} and the \textit{warm-hot} phase are evident.}
\end{figure*}

We have measured two sets of abundance ratios: (1) the abundance ratio of C, N, Ne, Mg and Si with respect to O and (2) $\alpha$/Fe (figure \ref{fig:nonsolar}). In our analysis, we adopt the solar composition model of \cite{Asplund2009}. The result does not change qualitatively if we use the relative solar abundances from \cite{Wilms2000} or \cite{Lodders2003}. This is not surprising, because the differences among the composition prescriptions are smaller than the statistical uncertainty of the parameters in our \texttt{PHASE} model. 

Please note that the abundance patterns in our observed gas phases do not necessarily reflect that from stellar nucleosynthetic yields. Metals in the ISM come from different sources e.g., winds and flares of stars in different phases of their age, core-collapse supernovae (CCSNe) and type Ia supernovae (SNIa), novae, globular clusters etc. Different sources produce and contribute metals to the ISM in different timescales, epochs and multiple generations/populations \citep{Nomoto2013}. The galactic wind, which carries metals from the ISM to the halo is not homogeneously mass-loaded for all metals \citep[e.g.,][]{Lopez2020}. The metal-enriched outflow is further mixed with the pre-existing lower metallicity gas in the halo. As a result, it is unlikely that the distribution of metals in the CGM would quantitatively trace/follow any stellar source of metals. Moreover, if the metals in the CGM are inhomogeneously mixed across phases \citep[{as has been observed in cooler phases of the CGM traced by \civn};][]{Schaye2007}, our measured chemical composition along a single sightline might be different from the global average. Nonetheless, we try to draw some qualitative inferences from the abundance ratios of metals obtained from our model. 

\subsection{\textit{Warm-hot} phase} \label{sec:warmhot}
The presence of the \textit{warm-hot} phase with log T(K)$\approx 6$ is known from X-ray absorption line studies \citep{Gupta2012,Gupta2014,Nicastro2016a,Nicastro2016b,Gupta2017,Nevalainen2017,Gatuzz2018,Das2019a}. The spatial distribution of the \textit{warm-hot} phase indicates a significant contribution from the Galactic halo beyond the disk \citep{Nicastro2016a,Nicastro2016b,Gatuzz2018}. Focusing on the sightline toward Mrk\,421, our estimation of the temperature and the hydrogen column density of the \textit{warm-hot} phase are consistent with those of \cite{Gatuzz2018}, who assumed a solar metallicity and solar chemical composition {in their ionization equilibrium model \texttt{IONeq}}. Their three-temperature model is equivalent with \texttt{tbabs} and one \texttt{PHASE} component in our model, containing the cold/cool ISM and the \textit{warm-hot} CGM. 

The non-thermal line broadening is consistent with theoretical simulations predicting that the highly ionized CGM is turbulent \citep{Rennehan2019,Bennett2020}, 
and that other non-thermal sources of motion such as magnetic fields and cosmic-rays also affect the velocity dispersion \citep{Voort2021,Ji2020}. The ratio of the non-thermal to the thermal broadening (see \S\ref{sec:results}) indicates the equipartition among the thermal and the non-thermal sources, or the dominance of the non-thermal source if only one kind of non-thermal source is present. 

\paragraph{Abundance ratios} \label{sec:abundT2}
Mg entirely comes from CCSNe \citep{Nomoto2013}, therefore the large value of [Mg/O] (and consequently super-solar Mg/Fe) may indicate CCSNe enrichment. However, a super-solar [O/Fe] also is expected in that case, but is not observed. This inconsistency may be resolved as follows. The higher emissivity of oxygen compared to Mg at $>10^6$\,K \citep{Bertone2013} implies that oxygen can cool more efficiently. In an inhomogeneously mixed CGM oxygen can transit to cooler phases without affecting the temperature of {other} elements, resulting in apparent deficit of oxygen. This has been observed {in transition temperature phase of the CGM traced by \oiv and \ov} \citep{Nevalainen2017}, and has been {indirectly suggested by cosmological simulations \citep{Ford2016,Grand2019}}. 

$\alpha$-enhancement in the \textit{warm-hot} Galactic CGM has been observed in emission studies \citep{Nakashima2018}, though with poor constraints and chances of confusion with foreground emission. In external galaxies, the $\alpha$-enhancement has been observed along the minor axes and in extra-planar regions within 30\,kpc of the galactic disks; this suggests that the $\alpha$-enhancement is associated with the galactic wind or fountain \citep{Strickland2004a,Yamasaki2009,Hodges-Kluck2020,Lopez2020}.  The $\alpha$-enhancement along our Mrk\,421 sightline, with $l=179.83^\circ$ indicates several possibilities:  

1) {Active galactic outflow resulting from CCSNe feedback may exist toward/around our line-of-sight. Alternatively, the metals from past nuclear outflows, as indicated by the Fermi bubble and the eROSITA bubble \citep{Su2010,Predehl2020}, may have spread far from the minor axis of the Milky\,Way. }

2) Iron is depleted onto interstellar dust that forms within 300-600 days after CCSNe explosions \citep{Todini2001}, leading to iron deficit in outflows. This has been found previously in the X-ray absorption spectra of Galactic low-mass X-ray binaries \citep{Pinto2013}. Alternatively, iron may be depleted onto extraplanar/circumgalactic dust. This process is stable, because there is a low chance of evaporation due to the low-density (10$^{-4}-10^{-3}$ cm$^{-3}$) environment \citep{Tielens1994}. This is substantiated by a larger fraction of CGM metals (42\%) in solid phase compared to that in the ISM ($\sim$30\%) in external galaxies \citep{Howk1999,Menard2010,Peeples2014}. 
\\ The depletion of iron onto dust is linked with outflow activities; therefore we cannot determine the relative importance of these two possibilities separately.

\subsection{\textit{Hot} phase} 
The strong \sixiv and \nex lines detected in the high S/N spectra have led us to discover the \textit{hot} (T$_3$) component.
Our observations by themselves cannot determine if the \textit{hot} phase resides in the Galactic ISM, CGM, or in the Local Group medium. We investigate different possible sources of the observed \textit{hot} phase to make a reasonable case. 

A) Our sightline is at high latitude ($b= 65.03^\circ$), passing through a small portion of the Galactic disk. Therefore, the chance of the \textit{hot} phase to come exclusively from the disk is small. The sightline is anti-center ($l = 179.83^\circ$), so there is no contribution from the X-ray shell around the Fermi bubble, North Polar Spur \citep{Kataoka2018} or the eROSITA bubble \citep{Predehl2020}. Moreover, dense structures such as supernovae remnants (SNR) or superbubbles made by expanding and merging SNR should have detectable emission measure due to their high density. But, the X-ray emission studies along/close to our sightline have not found any such structure yet \citep{Yao2007a,Sakai2014,Gupta2014}. This implies that the contribution of SNR or superbubbles to the \textit{hot} gas, if any, is unlikely to be significant.  

B) We investigate whether the \textit{hot} gas is from the Local Group. Assuming the Local Group mass $= 6.4\times 10^{12}$ M$_\odot$ \citep{Peebles1990}, and using the $T_X-M$ relation for galaxy clusters: $T_X \propto M^{{2/3}}$, the temperature of the Local Group is $\approx 10^{6.69-6.91}$\,K. This is smaller than T$_3$ by more than 3$\sigma$ (see figure \ref{fig:contour}, bottom panel). Moreover, our sightline ($l=179.83^\circ, b=65.03^\circ$) is away from M31 ($l=121.17^\circ, b=-21.57^\circ$) and the barycenter of the Local Group. Therefore, the contribution of the Local Group to the \textit{hot} phase, if any, is unlikely to be significant. The CGM of MW-like halos of M$_{200} \leqslant 10^{12}$ M$_\odot$ can be extended beyond the virial radius (R$_{200}$) if the thermal feedback buoyantly rises to the outer CGM and moves the baryons out \citep{Oppenheimer2018}. Therefore, if the observed \textit{hot} gas is extended beyond R$_{200}$, it is just a matter of nomenclature whether it should be called the CGM or the Local Group medium. In either case, the gas would be out of thermal and hydrostatic equilibrium due to its higher temperature than the virial temperature of the Milky\,Way CGM and the Local Group.

C) The \textit{hot} phase might originate from active outflows driven by stellar winds and supernovae (SNe), expanding into the halo. The X-ray emission analysis of the nuclear outflow from M82 indicates the co-existence of a \textit{hot} phase with the \textit{warm-hot}  phase \citep{Ranalli2008,Lopez2020}. Using the physical parameters of the outflow derived in those analyses, we calculate the approximate column density of a similar outflow if it is observed from the location in the disk. The resulting column density at solar metallicity $\approx2.8\times10^{20} cm^{-2}$ is consistent with that of the \textit{hot} phase within 2$\sigma$ (figure \ref{fig:contour}, panel g). While Milky\,Way is not a starburst galaxy unlike M82, the consistency in the column densities indicates that the possibility of an outflow cannot be ruled out. Because the column density is degenerate with the density and the path length, a different combination of these parameters (e.g., an outflow with lower density and larger volume) might produce the same column density. A careful search for any emission signature around $\approx$1\,keV toward/close to our sightline would test this scenario (Bhattacharyya \textit{et al.} 2021, in prep.). 

There have been hints of \textit{hot} gas in the halo of the Milky\,Way in emission away from the Galactic center \citep{Yoshino2009,Mitsuishi2012,Henley2013,Nakashima2018,Gupta2021}. It shows that the \textit{hot} gas can be present without any relation to the nuclear activity of the Galaxy. However, the emission from the \textit{hot} gas may be confused with the charge-exchange emission and/or the overabundance of metal(s); this is not an issue in absorption analyses. The unambiguous presence of the \textit{hot} gas in absorption shows that the \textit{hot} gas is truly widespread. 


Similar to the \textit{warm-hot} phase, the line broadening in the \textit{hot} phase indicates significant contribution of non-thermal sources (see \S\ref{sec:warmhot}). The ratio of the non-thermal to the thermal broadening in the \textit{warm-hot} and \textit{hot} phases are similar (see \S\ref{sec:results}); this indicates that these two phases are related to/influenced by each other.  

The $\alpha$-enhancement, and the values of [Ne/O] and [Si/O] are consistent with CCSNe enrichment \citep{Nomoto2013}. It favors the possibility of an outflow rather than the pre-existing gas in the halo (or beyond). This was also observed in the 10$^7$\,K hot phase along 1ES\,1553+113 \citep{Das2019a}. 

Our temperature estimates are subject to the assumption that the gas is in collisional ionization equilibrium. However, the \textit{hot} gas is not at the virial temperature of Milky\,Way or the Local Group, and is likely related to galactic outflows. Therefore, it might not be in an equilibrium state. 
If the gas producing the \sixiv and \nex absorption lines is photo-ionized, it would affect the derived physical quantities of the \textit{hot} phase. The effects of the non-equilibrium collisional ionization and  photo-ionization are degenerate in metal-enriched cooling gas \citep{Oppenheimer2013}. Therefore, estimating their individual contribution to the \textit{hot} phase, if any, is non-trivial and beyond the scope of this paper. 


\subsection{\textit{Warm} phase} 
The \cv K$\alpha$ and K$\beta$ lines have led to the detection and characterization of the \textit{warm} (T$_1$) component, for the first time in X-ray absorption\footnote{There was a suggestive evidence of this phase in \cite{Williams2005}, but our measurement is more unambiguous due to better quality of data, larger spectral coverage and self-consistent ionization modeling to simultaneously fit all phases}. This phase is usually observed in UV probed by primarily \ovi (and sometimes, \neviiin). Similar to the case of \textit{hot} phase, we cannot determine the location of the \textit{warm} phase along our line-of-sight, and cannot say whether it is co-spatial with the \textit{warm-hot} phase.

\paragraph{Carbon lines}
The \textit{warm} phase produces \civ in addition to \cvn, and our \texttt{PHASE} model predicts N(\civn)$_X$ $= 8.97^{+0.76}_{-1.42} \times 10^{12}$ cm$^{-2}$. The N(\civn) measured in high velocity clouds (HVC : $|v_{LSR}| = 100-500$\,km s$^{-1}$) using UV absorption along the same sightline is N(\civn)$_{UV}$ = $5.89\pm1.22 \times 10^{12}$ cm$^{-2}$ \citep{Richter2017}. These two are consistent with each other within $2\sigma$, but our model predicts N(\civn)$_{X}$-N(\civn)$_{UV}$ $= 3.08^{+1.44}_{-1.87} \times 10^{12}$ cm$^{-2}$ of excess N(\civn). We cannot apply any velocity cut in our X-ray analysis; therefore this excess absorption, if present, is likely to originate in low/intermediate velocity absorbers in the disk/halo. The UV \civ HVC observations suggests that the \textit{warm} phase resides in the halo rather than in the disk. It also indicates that the detected \civ in UV along this sightline is unlikely from a cool ($10^4<$T$<10^5$\,K) phase co-probed by \ciin, as was assumed in \cite{Richter2017}. This reinstates the reason to study multiple metal lines in X-ray absorption and also shows the power of multi-wavelength analysis. 

\paragraph{Oxygen lines}\label{par:ov}
The \textit{warm} phase also produces \ov and \ovi along with \ovii (see \S\ref{sec:results}). Our \texttt{PHASE} model predicts N(\ovin) $= 1.82^{+0.14}_{-0.27} \times 10^{14}$\,cm$^{-2}$. But N(\ovin) measured in UV absorption along the same sightline is  $2.66-3.15 \times 10^{14}$\,cm$^{-2}$ \citep{Williams2005,Yao2007b}, factor of $\approx2$ larger than our prediction. That means the \textit{warm} phase does not account for all the \ovi detected in UV. Assuming that the absorption feature at 22.03\AA~in our spectrum is due to \ovi K$\alpha$ line only, the total N(\ovin) is $2.15^{+0.53}_{-0.50} \times 10^{15}$\,cm$^{-2}$, larger than that from UV analysis by an order of magnitude. This is due to the blending of \oii~K$\beta$ line at 22.04\AA, which accounts for most of the absorption \citep{Mathur2017}. To test if the excess \ovi in UV is due to photo-ionization, we refit the X-ray spectrum with the photo-ionization parameter \textit{U} of the \textit{warm} phase frozen\footnote{because the ionization effect of the temperature and the photo-ionization are degenerate with each other \citep{Oppenheimer2013}, we freeze \textit{U} at different values instead of allowing both the temperature and \textit{U} to vary} at 10$^{-3}$, 10$^{-2}$ and 10$^{-1}$. The best-fit models have worse $\chi^2/dof$ than that of the best-fit model with \textit{U}=10$^{-4}$, indicating that the latter describes the spectrum better. Moreover, none of the models with higher value of \textit{U} can account for the excess \ovi measured in UV. This suggests that something other than photo-ionization might be affecting the \textit{warm} phase, making it deviate from collisional ionization equilibrium. This is consistent with previous findings \citep{Lochhaas2019}, where modeling the tracer ions of this phase in the Galactic halo indicated the effects of radiative cooling or non-thermal sources e.g., turbulent mixing.

The upper limit of the non-thermal broadening in the \textit{warm} phase ({figure} \ref{fig:contour}, panel b) indicates that the non-thermal sources are not as strong as they are in the hotter phases, if at all present. While the detailed effect of each source of non-thermal motions (turbulence, magnetic fields, cosmic ray) on the physical properties of the CGM is different, all of them enhance the cooling of the hot halo gas and generate cooler and denser phases to balance the total (thermal and non-thermal) pressure against gravity \citep{Bennett2020,Ji2020,Voort2021}. It implies that the non-thermal broadening in the \textit{warm-hot} and the \textit{hot} phases is a natural indicator of coexistent phase(s) at lower temperature. This is consistent with the picture of the \textit{warm} gas having formed from the cooling of the \textit{warm-hot} phase.

{
If the \textit{warm} phase is clumpy/cloudy instead of diffuse, the phase structure might be more complex. Instead of a uniform \textit{warm} phase, individual clouds may contain different combination of metal ions at different temperatures. In that case, the average [C/O] would not be informative and physically meaningful. The super-solar [C/O] might be due to the deficit of oxygen because of its preferential cooling and inhomogeneous mixing of metals while cooling from hotter phases, as discussed in the case of the \textit{warm-hot} phase (see \S\ref{sec:abundT2}).}

\subsection{An emerging picture}\label{sec:emergingpic}
The combined analysis of X-ray emission and absorption toward 1ES\,1553+113 indicated the co-existence of at least three temperatures between $\approx 10^5 - 10^8$\,K  \citep{Das2019a,Das2019c}. Our results in this paper are consistent with this picture. The temperature of the phases span more than two orders of magnitude; therefore it is unlikely to be a result of local temperature fluctuation or temperature dispersion in a single phase as is often assumed in theoretical simulations \citep[e.g.,][]{ Faerman2017,LiMiao2020}. 

In cosmological zoom-in simulations and idealized simulations of individual galaxies \citep[see their comparative analysis in][]{Fielding2020}, the CGM of a MW-like galaxy is predicted to be a diffuse, volume-filling, low-density, massive \textit{warm-hot} medium that is cooling due to thermal and hydrodynamic instability, resulting in a structure distributed over a large range of density and temperature. The lower ionization phases at T=$10^{4-5}$\,K and predominantly neutral phase at T$<10^4$\,K are visualized as dense, clumpy structures that are cooled from and embedded in the hotter medium, with the intermediate ionization \textit{warm} phase at T=$10^{5-6}$\,K at the interface \citep{Marasco2013,Voit2015,Armillotta2017,Strawn2021}. 

Most of the observational studies have focused on the oxygen lines (\ovin, \oviin, \oviiin) in the $>10^5$\,K phases. However, the 10$^7$\,K gas is not detectable with oxygen-only analysis because oxygen is fully ionized by that temperature. Detection of other heavier ions with weaker lines is required to uncover the hot gas, which is difficult to do with low SNRE spectra. Therefore, the apparent lack of the 10$^7$\,K gas in observations of the Milky Way and MW-like galaxies might be an observational bias. Existing simulations \citep[e.g.,][]{Ford2016,Nelson2018,Wijers2019} have successfully reproduced the observed lines of oxygen. However, most of the papers do not predict the physical properties of the \textit{hot} gas, such as the density and pathlength. Recently, a realistic outflow model including the interaction between the outflow and the halo of a MW-like galaxy has been able to reproduce the 10$^7$\,K gas for SFR = 10 \msun yr$^{-1}$, but for a MW-like SFR the 10$^7$\,K gas is still not abundant  \citep{Vijayan2021}. This indicates the need for improved modeling of thermal feedback of galactic outflows and/or non-thermal heating by turbulent dissipation, magnetic field reconnection, and cosmic-rays \citep{Reynolds1999,Jana2018}, and high spatial resolution to resolve the complex phase structure of the CGM of a MW-like galaxy and consistently produce the 10$^7$K gas.  

The temperature we calculate for the diffuse medium is the density-weighted radial average of the temperature: $T = \frac{\int n(l) T(l) dl}{\int n(l) dl}$, {where $T(l)$ and $n(l)$ are the temperature and the hydrogen number density distribution along our line-of-sight, respectively}. For a clumpy medium, it is the average of individual clouds where the variation in temperature is more like a fluctuation rather than a smooth distribution. Additionally, there might be a metallicity gradient. We do not have sufficient information to estimate the parameters of the density/temperature/metallicity distribution and the path length independently. 

The temperature estimate is generally biased by the most abundant ion in that phase, e.g., \cv and \ovii in the \textit{warm} and the \textit{warm-hot} phases, respectively. Different metal ions in a given phase need not be co-spatial; they may occupy a higher/lower temperature region according to their ionization fraction. This would underestimate the expected ionization fraction of the less abundant element at the best-fitted temperature, and would overestimate its abundance ratio with respect to the most abundant element. This may partially explain the high [C/O] and [Mg/O] in the \textit{warm} and \textit{warm-hot} phases, respectively. 

The three discrete phases in our model is a simplified picture of the $>10^5$\,K CGM. Due to the limited baseline in the observed spectra, our results are restricted to a certain range of temperature. The absolute metallicities, the morphology, and the density distribution of the observed system are degenerate with the temperature. This makes the estimation of a more realistic and meaningful phase structure along a single sightline extremely challenging. For theoretical purpose, the best-fit value and the statistical uncertainty of a parameter (see table \ref{tab:best-fit}) can be used as the proxy of the peak and the dispersion of its (e.g., log-normal) distribution, respectively. However, the contours in figure \ref{fig:contour} show that the uncertainties are sensitive to the type of correlation between different parameters, so this approximation should be used with caution. We need to model a larger sample of sightlines across the sky to get a better handle on the parameters and their global dispersion; this would be the part of a future endeavor. 

From the above discussion, we infer the following picture of the Milky Way CGM. The \textit{hot} gas is the metal-enriched, dense, nearby phase tracing galactic wind expanding into the halo. In that case, the actual column density of the \textit{hot} phase would be smaller because the column density is inversely proportional to metallicity, N(H)$_Z$ = N(H)$_\odot \times (Z/Z_\odot)^{-1}$. There are additional non-thermal sources of heating in the \textit{hot} phase. The \textit{warm-hot} phase is the quasi-static phase in the halo where non-thermality is induced by the outflow. The \textit{warm} phase is tracing clouds or the outer layer of cooler clouds embedded in the \textit{hot} and the \textit{warm-hot} phases, outflowing and/or inflowing. This, albeit under a lot of assumptions, {could be} the emerging picture of the Galactic CGM.  

\subsection{Missing Galactic baryons and metals}
Studies involving absorption analysis indicate that the extended \textit{warm-hot} CGM of the Milky Way likely accounts for all the missing Galactic baryons \citep{Gupta2012,Nicastro2016b}. However, emission analysis shows that the \textit{warm-hot} phase is dominated by the disk whose mass is insufficient to account for the missing Galactic baryons \citep{Kaaret2020}. This suggests that the emitting and the absorbing \textit{warm-hot} gas do not sample the same space in the Galaxy, and their masses should be estimated separately. Secondly, as discussed in \S\ref{sec:phase} (also see table \ref{tab:lineprofiles}), the column density (and the mass) would be overestimated if it is obtained by assuming a single phase. Therefore, the multi-phase ionization modeling is crucial to correctly estimate the column density, metallic mass, and the total baryonic mass (for a given metallicity) of the \textit{warm-hot} phase. 

The baryonic and the metallic mass of the \textit{hot} phase has not been estimated yet. Including Mrk\,421, the 10$^7$\,K gas has been detected along 11 sightlines in X-ray absorption and/or emission analyses across the sky \citep{Yoshino2009,Mitsuishi2012,Henley2013,Das2019a,Das2019c,Gupta2021}. There are no reported non-detections of this phase where it has been searched for. As a result, we do not know the covering fraction of the \textit{hot} phase. Its mass would be sensitive to its metallicity and density distribution, {and its path length, i.e.,} whether it is confined in the extraplanar region or is extended {out to a significant fraction of the virial radius}. Nonetheless, its column density and emission measure along the 11 sightlines suggest that it will contribute a non-negligible amount to the Galactic budget of baryons and metals.

\subsection{Dispersion measure}
The dispersion measure (DM) of fast radio bursts (FRBs) is a promising tool to probe the ionized baryons in the intergalactic medium (IGM) and the CGM of external galaxies  \citep{Li2019,Prochaska2019}. For any extragalactic calculation, it is necessary to remove the Galactic contribution, $DM_{Gal}$ from the total observed DM toward an FRB. The current estimate of $DM_{Gal}$ based on X-ray absorption and emission studies \citep{Prochaska2019,Yamasaki2020,Das2021a} include the contribution of the \textit{warm-hot} phase only. The \textit{hot} phase will increase the $DM_{Gal}$ along and around the directions where this phase is present. This will significantly impact the extragalactic studies using DM of FRBs. As mentioned in \S\ref{sec:emergingpic}, estimating the physical properties of the \textit{hot} phase and its contribution to the $DM_{Gal}$ across the sky would be part of a future endeavor.  

\subsection{Intervening absorption?}\label{sec:igm}
The low-redshift intergalactic medium (IGM) is believed to be warm-hot, detectable through absorption lines of highly ionized metals \citep[and references therein]{Nicastro2002,Mathur2003}. The pathlength for detecting intervening absorption lines toward Mrk\,421 at $z$=0.031 is small, but the blazar is X-ray bright, offering an interesting opportunity. \cite{Nicastro2005} reported the detection of two intervening absorption systems, possibly from the warm-hot IGM (WHIM), at $z_1=0.011\pm 0.001$ and $z_2=0.026\pm 0.001$. Their best $z_1$ line, detected at $3.8\sigma$, is \ovii $K\alpha$ with EW$=3.0^{+0.9}_{-0.8}$\,m\AA. We do not detect this line in the extremely high S/N spectrum reported here; the $3\sigma$ upper limit is $3.1$\,m\AA. The best $z_2$ line in \cite{Nicastro2005} is the \nvii $K\alpha$ line at $3.1\sigma$ significance with EW$=3.4\pm 1.1$\,m\AA. We do not detect this line either, with the $3\sigma$ upper limit of $1.9$\,m\AA. Also, we do not detect any other $z_2$ line. However, we detect the \oviii $K\alpha$ line at $z_1$ with EW$=2.2\pm0.6$\,m\AA, which is consistent with the upper limit of $4.1$m\AA~in \cite{Nicastro2005}. {We tentatively detect a \sixiv K$\alpha$ line at $z_1$ with EW$=0.9\pm0.3$\,m\AA, which was not reported in \cite{Nicastro2005}. If true, the non-detection of \ovii and the detection of \oviii and \sixiv might indicate a hotter WHIM}. Similarly, we detect the \oviii $K\alpha$ line at $z_2$ with EW$=2.2\pm0.6$\,m\AA, which is consistent with the upper limit of $1.8$m\AA~within 1$\sigma$ \citep{Nicastro2005}. Thus, the detection of intervening absorption in the Mrk\,421 sightline remains inconclusive. 

\section{Conclusion}\label{sec:summary}
\noindent In this paper, we have studied the z=0 absorber(s) in the deep X-ray grating spectra toward Mrk\,421, covering a large wavelength range of 5-43\AA. Our analysis has produced a wealth of important results presented below:

$\bullet$ We {detect} a \textit{hot} $\approx10^{7.5}$\,K gas phase probed with \sixiv and \nexn. This phase coexists with the \textit{warm-hot} 10$^{6.2}$\,K and the \textit{warm} 10$^{5.5}$\,K CGM, along this sightline. This is the first time that a three-phase highly ionized CGM has been detected and characterized in detail by absorption analysis. 

$\bullet$ The hotter phases have higher hydrogen column density than the \textit{warm} phase, for solar metallicity of oxygen. This indicates that the hotter phases might have higher metallicity/density/path length than the \textit{warm} phase.   

$\bullet$ For the first time, we have been able to constrain the non-thermal line broadening of any highly ionized CGM phase. The \textit{warm-hot} and the \textit{hot} phases of the CGM have non-zero non-thermal line broadening. The non-thermal broadening dominates the broadening of the metal lines. 
    
$\bullet$ The abundance ratios of metals in the CGM are inconsistent with a solar-like chemical composition: 

$\ast$ The \textit{hot} CGM is significantly $\alpha-$enhanced, likely due to CCSNe enrichment and/or Fe depletion onto dust. [Ne/O] and [Si/O] are super-solar in the \textit{hot} phase, and is consistent with CCSNe enrichment

$\ast$ [Mg/Fe] (and [Mg/O]) are significantly super-solar in the \textit{warm-hot} phase, {suggesting} evidence of CCSNe enrichment and preferential cooling of oxygen to lower temperature  

$\ast$ [C/O] is super-solar in the \textit{warm} phase. This could be due to inhomogeneous mixing in clouds while cooling from hotter phases or a complex phase structure of clouds in the  10$^{5-6}$\,K range.
These results provide insights into the thermal history, non-thermal effects, chemical enrichment, mixing and depletion in the circumgalactic medium, and provide important inputs to theories of galaxy evolution. 

It is necessary to extend such deep X-ray absorption analysis to many other sightlines to search for and characterize the temperature, column density, kinematics and chemical composition of the highly ionized CGM, using multiple tracer elements along with oxygen. At present, the archival data of \chandra and \xmm can be very useful in this regard. On a longer timescale, upcoming missions like \textit{XRISM, Arcus and Athena} and the concept mission like \textit{Lynx} in the next decade and beyond will offer an outstanding opportunity to observe the highly ionized diffused medium in unprecedented detail. Semi-analytical/ high-resolution numerical simulations should also look for multi-phase highly ionized CGM to explain the observations. This will bring us closer to understanding the co-evolution of the Galaxy and its CGM.

\section*{acknowledgments}
\textit{
Support for this work was provided by the National Aeronautics and Space Administration through Chandra Award Numbers GO7-18129X and AR0-21016X to S.M. issued by the Chandra X-ray Center, which is operated by the Smithsonian Astrophysical Observatory for and on behalf of the National Aeronautics Space Administration under contract NAS8-03060. S.M. is also grateful for the NASA grant NNX16AF49G. A.G. gratefully acknowledges support through the NASA
ADAP 80NSSC18K0419. Y.K. acknowledges support from DGAPA-PAPIIT 106518.}

\facilities{\chandra}
\software{CIAO 4.13 \citep{Fruscione2006} , HeaSoft v6.17 \citep{Drake2005}, XSPEC 12.11.1 \citep{Arnaud1999}, NumPy 1.19.5 \citep{Dubois1996}, Matplotlib 3.3.4 \citep{Hunter2007}}

\appendix 
\restartappendixnumbering
\section{Continuum fitting of Mrk\,421} \label{app:continuum}

{We start the analysis by fitting the global continuum. It is modeled as a variable index powerlaw, absorbed by the cold Galactic ISM (\texttt{eplogpar*tbabs}). The commonly used models such as powerlaw, broken powerlaw and blackbody or their combination do not fit the continuum well ($\chi^2/dof \gg 1$). Therefore, we include \texttt{eplogpar}, a powerlaw with an index which varies with energy as a log parabola. This function is used to explain sources of electron acceleration in BL Lac objects \citep{Tramacere2007,Goswami2020}. The source continuum is absorbed by the Galactic ISM; we describe that with the Tuebingen-Boulder ISM absorption model \citep{Wilms2000}. This model calculates the cross section for X-ray absorption due to the gas-phase, the grain-phase and the molecules in the ISM. It assumes solar metallicity and solar-like chemical composition, and includes high resolution structures for the K-edges of oxygen, neon and the L-edges of iron. }

{To fit the residual curvatures in the global continuum which likely arise due to calibration uncertainties, we include broad, resolvable Gaussian profiles with both positive and negative normalization as per need, as has been done previously \citep{Nicastro2002,Williams2005}. We allow the wavelength, width and the normalization of these profiles to vary.}

{There are several $>3 \sigma$ narrow residual decrements not at the locations of $z=0$ He-like and H-like metals. We model them with unresolved Gaussian absorption profiles. We list the wavelength, equivalent width and possible identification of these lines in Table \ref{tab:extralines} and these are discussed further below in Appendix \ref{app:otherlines}. }

{We define the pseudo-continuum as defined above and then look for the $z=0$ absorption lines from the MW CGM. For our model-independent study, it is sufficient to fit the local continuum around the strongest transitions of He-like and the H-like metal ions. But, our ionization modeling (see \S\ref{sec:phase}) includes weaker transitions of these ions in addition to the stronger ones, which are spread across the global continuum. These transitions are not individually detectable, but the constraint on their strength makes the estimation of the physical parameters more precise and self-consistent than the model-independent analysis.
This justifies the detailed fitting of the global continuum before searching for the He-like and the H-like  metal ions.}

\section{Absorption lines not from the Milky Way CGM} \label{app:otherlines}

{As noted in Appendix \ref{app:continuum}, there were other absorption lines in the spectrum which were not at the location of the $z=0$ He-like and H-like metal lines. These are listed in Table \ref{tab:extralines} and we discuss them here.}

{Some of these ``other" lines are from the putative intervening absorbers at $z=0.011$ (see discussion in \S\ref{sec:igm}). The $10.24$ \AA~line could be a blend of \nex K$\beta$ \citep[10.24 \AA,][]{Erickson1977} and $2s\rightarrow 3d$ transition of \nixxiv \citep[10.28 \AA,][]{Verner1995}; this is the first astronomical identification of this line to our knowledge.}

\begin{figure}[t]
    \centering
    \includegraphics[scale=0.5]{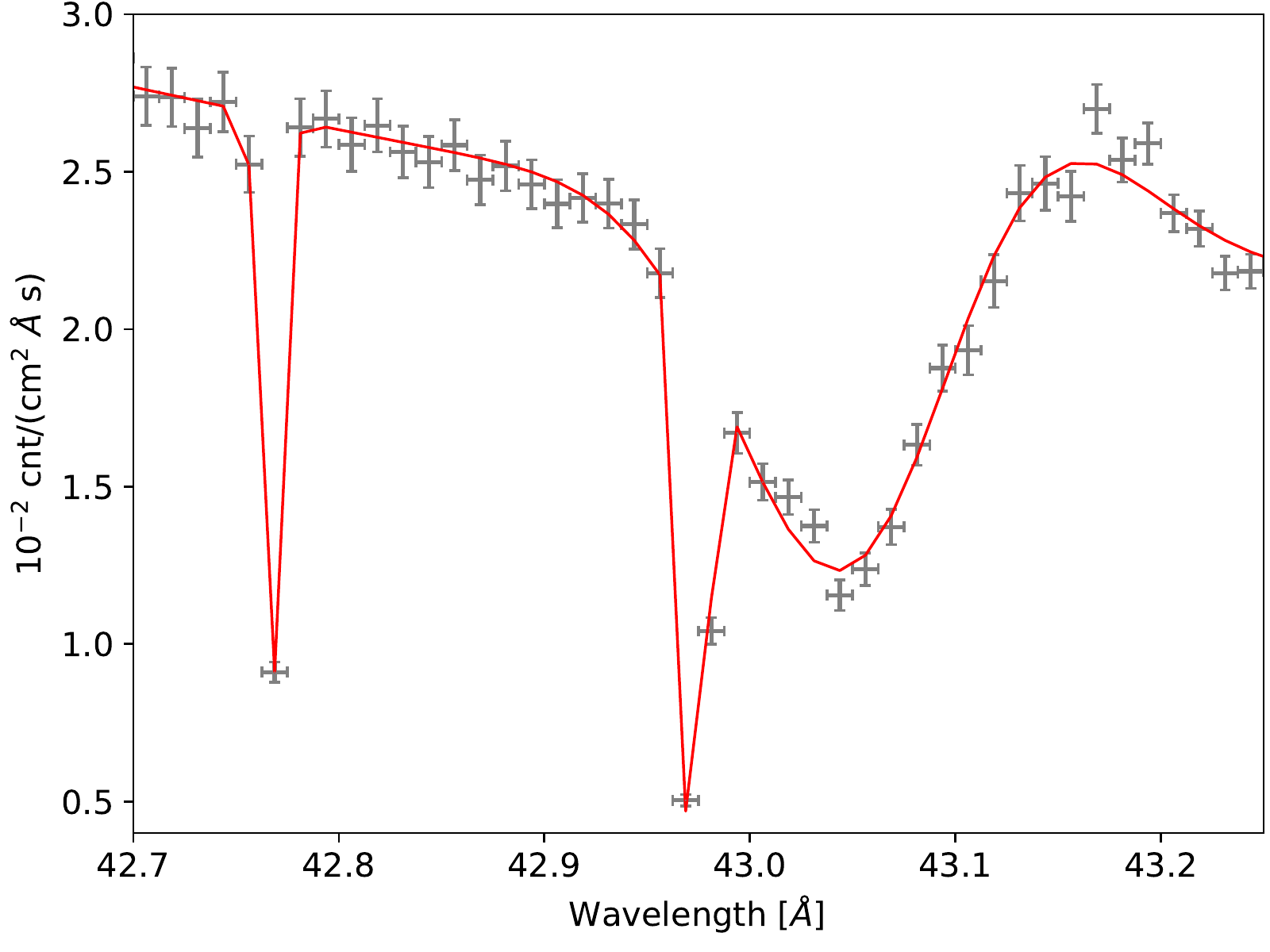}
    \caption{\cii K$\alpha$ triplet toward Mrk\,421.}
    \label{fig:CII}
\end{figure}

We detect two blends of \ovi and \oii~K-shell transitions likely dominated by \oii:  \oii~K$\beta$ \citep[22.04 \AA,][]{Bizau2015} and \ovi K$\alpha$ \citep[22.03 \AA,][see further discussion in \S\ref{par:ov}]{McLaughlin2017}; and \oii~K$\beta$ \citep[22.30 \AA,][]{Bizau2015} and \ovi K$\alpha$ \citep[22.30 \AA,][]{Liang2011}. We detect \nin--\niii K$\alpha$ lines at 30.53--31.28 \AA~\citep{Garcia2009,Gatuzz2021N}. The absorption at 40.04 and 40.34 \AA~could be weaker K-shell transitions of \cii \citep{Hasoglu2010,Gatuzz2018C}. The strongest transition of \cii K$\alpha$ triplet between 42.7-43.2 \AA~are shown in figure \ref{fig:CII}. The detection of strong \ciin, \nin--\niii  and \oii~lines indicate that these transitions are underestimated by \texttt{tbabs}, suggesting super-solar carbon, nitrogen and oxygen in the Galactic cold/cool ISM toward Mrk\,421. Other equivalent ISM models in \texttt{XSPEC} (e.g., \texttt{wabs,phabs,ISMabs} etc.) do not include all the low ions of carbon, nitrogen and oxygen in full capacity either. Therefore, replacing the \texttt{tbabs} model does not improve the situation. Precise information about these ion transitions, e.g., the wavelength, photo-absorption cross section etc. are still a topic of active research, and often, well-understood astrophysical sources are used for benchmarking.  Because our science interest is not focused on the physical properties of the cold/cool ISM, we do not discuss this further in the main text.

\begin{deluxetable}{ccc}\label{tab:extralines}
\tablecaption{{Other absorption lines with $\gtrapprox 3 \sigma$ detection}}
\tablehead{
\colhead{Wavelength (\AA)} & \colhead{EW  (m\AA)} & \colhead{Identification}
      }
\startdata
    \multicolumn{3}{c}{ACIS-MEG}\\
    \hline
    6.26   & 0.88$\pm0.29$   &  \sixivn$_{z=0.011}$? \\
    \hline
    \multicolumn{3}{c}{ACIS-LETG}\\
    \hline
    10.24 & 1.67$^{+0.48}_{-0.22}$ & \nexn+\nixxiv? \\
    19.15 & 2.17$\pm0.61$ & \oviiin$_{z=0.011}$ \\
    19.51 & 2.18$\pm0.64$ & \oviiin$_{z=0.026}$  \\
    22.02 & 3.79$\pm0.87$ & \oii+\ovin \\
    22.29 & 3.07$^{+0.97}_{-0.85}$ & \oii+\ovin \\
    \hline
    \multicolumn{3}{c}{HRC-LETG}\\
    \hline
    25.20 & 2.78$^{+0.43}_{-0.84}$ & unidentified/spurious$\dagger$ \\
    26.94 & 2.72$\pm0.65$ & unidentified  \\
    29.04 & 2.57$^{+0.67}_{-0.71}$ & unidentified/spurious$\dagger$ \\
    30.15 & 2.15$\pm0.76$ & unidentified/spurious \\
    30.53 & 2.22$^{+0.81}_{-0.72}$ & \niii K$\alpha$ \\
    30.92 & 7.02$^{+0.26}_{-1.68}$ & \nii K$\alpha$ \\
    31.00 & 5.15$^{+0.74}_{-0.65}$ & \nii K$\alpha$ \\
    31.28 & 3.78$^{+0.61}_{-0.82}$ & \ni K$\alpha$ \\     
    40.04 & 5.20$^{+0.67}_{-1.29}$ & \cii?$^b$ \\  
    40.34 & 7.56$^{+1.36}_{-1.00}$ & \cii?$^b$ \\  
\enddata

Note-\footnotesize {The uncertainties in the wavelengths are same as the resolution elements, i.e., 12.5 m\AA~for ACIS-MEG, 25 m\AA~for ACIS-LETG and HRC-LETG.  Unidentified lines are likely the transitions of lower ionization states of  light elements (e.g., Si, S, Ar, Ca etc.) from the cold/cool Galactic ISM (see \href{http://www.atomdb.org/}{AtomDB}). The equivalent width of spurious lines are $>2\sigma$ different between positive and negative order spectra, implying that they are likely caused by bad pixels. }

$\dagger$\footnotesize {K-shell transitions of \nv have been predicted at these wavelengths \citep{Garcia2009}. But, as the strongest transition of \nv K$\alpha$ at 29.42 \AA~is not detected in our spectra (Figure \ref{fig:spectrum}, bottom), these are unlikely to be \nv lines. }



\end{deluxetable}

{This leaves us with four lines that are unidentified and/or spurious. To further test these  possibilities, we looked for the line signatures in both positive and negative orders of the gratings. The line at $26.94$ \AA~was detected in both the orders, therefore it is unlikely to be spurious; we report this line as ``unidentified" in Table \ref{tab:extralines}. Several lines reported in Table \ref{tab:extralines} were identified for the first time in the past few years. This is an emerging field, therefore it is quite possible that the unidentified lines will be identified in the years to come. However, we cannot rule out the possibility that the remaining three lines are truly spurious. The HRC-LETG spectrum has 784 resolution elements. Therefore, we expect to detect 2 lines with $>3\sigma$ significance by chance, as observed at 25.2 \AA~and 29.04 \AA. The third unidentified/spurious line at 30.15 \AA~is detected with 2.82$\sigma$ significance, but we report it for the sake of completeness. Despite being a $<3 \sigma$ decrement, we add the 30.15 \AA~line in our pseudo-continuum model to properly estimate the local continuum (a broad Gaussian between \nvi K$\alpha$ and \niii lines) around \nvi K$\alpha$ which is necessary for our science interest. We expect to detect 3 lines with $>2.82\sigma$ significance by chance, consistent with our observations.}

\textit{Data and model availability:} We have used public archival data in our analysis. The reprocessed and stacked data and the continuum model are available from the corresponding author upon request.  


\end{document}